\documentclass[12pt, draftclsnofoot, onecolumn]{IEEEtran}

\usepackage{cite,graphicx,amsmath,amssymb}
\usepackage{breqn}
\usepackage{subfigure}
\usepackage{fancyhdr}
\usepackage{mdwmath}
\usepackage{mdwtab}
\usepackage{balance}
\usepackage{xcolor}
\usepackage{bm}
\usepackage{amsthm}

\usepackage{multirow}
\usepackage{flafter}
\usepackage{caption}
\usepackage[english]{babel}

\usepackage{tcolorbox}
\usepackage{framed}
\usepackage{adjustbox}
\usepackage{blindtext}

\usepackage[normalem]{ulem}%uline

\usepackage[ruled,vlined]{algorithm2e}

\usepackage{epstopdf}

\usepackage{hyperref}
\hypersetup{
	colorlinks=true,
    allcolors = black
}

\usepackage{breqn} % automatic line break

\newtheorem{remark}{Remark}
\newtheorem{theorem}{Theorem}

\newtheorem{lemma}{Lemma}

\newtheorem{corollary}{Corollary}

\newtheorem{proposition}{Proposition}

\definecolor{coolgrey}{rgb}{0.55, 0.57, 0.67}
\definecolor{airforceblue}{rgb}{0.36, 0.54, 0.66}
\definecolor{americanrose}{rgb}{1.0, 0.01, 0.24}
\definecolor{amber}{rgb}{1.0, 0.49, 0.0}

\begin{document}

\title{Quantum Approximate Optimization Algorithm Based Maximum Likelihood  Detection }

%\vspace{-3.0em}
\author{
Jingjing Cui,~\IEEEmembership{Member, IEEE,}
Yifeng Xiong,~%~\IEEEmembership{Student Member, IEEE,} 
Soon Xin Ng, ~\IEEEmembership{Senior Member, IEEE} 
and Lajos Hanzo,~\IEEEmembership{Fellow, IEEE}
\vspace{-2.9em}

\thanks{J. Cui, Y. Xiong, S. Ng and L. Hanzo  are with the School of Electronics and Computer Science, University of Southampton, SO17 1BJ, Southampton (UK) (e-mail: \{jingj.cui, Yifeng.Xiong\}@soton.ac.uk and \{sxn,lh\}@ecs.soton.ac.uk).}

}

\maketitle

\begin{abstract}

Recent advances in quantum technologies  pave the way for   noisy intermediate-scale quantum (NISQ) devices, where  quantum approximation optimization algorithms (QAOAs) constitute  promising candidates for demonstrating tangible quantum advantages based on NISQ devices. In this paper,  we consider the maximum likelihood (ML) detection problem of binary symbols transmitted over a multiple-input and multiple-output (MIMO)  channel, where finding the optimal solution is exponentially hard using classical computers. Here, we apply the QAOA  for  the ML detection by encoding the problem  of interest  into a level-$p$ QAOA circuit having $2p$ variational parameters, which can be optimized by classical optimizers.  This level-$p$ QAOA circuit is constructed by  applying the prepared Hamiltonian to our problem and the  initial Hamiltonian alternately in $p$ consecutive rounds. More explicitly, we first encode the optimal solution of the ML detection problem  into the ground state of a problem  Hamiltonian.  Using the quantum adiabatic evolution technique, we provide both  analytical and numerical results for characterizing the evolution of the eigenvalues of the quantum system used for  ML detection. Then, for level-1 QAOA circuits, we derive the analytical expressions of the expectation values of the QAOA and discuss the complexity of the QAOA based ML detector. Explicitly,  we evaluate the computational complexity of the classical optimizer used and the storage requirement of  simulating the QAOA. Finally, we  evaluate the bit error rate (BER) of the QAOA based  ML detector and  compare it both to the classical ML  detector and to the classical minimum mean squared error (MMSE)  detector, demonstrating that the QAOA based  ML detector  is capable of approaching the performance of the classical ML detector. This paves the way for a host of large-scale classical optimization problems  to be solved by NISQ computers.

\end{abstract}
%%}
%%\vspace*{-0.4em}

\section{Introduction}
\label{sec:intro}
The evolution of social networking and  the ubiquitous wireless connectivity, coupled with the availability of low-cost yet powerful
computing devices  jointly shape the  next generation of wireless systems.   Integrated ground-air-space
(IGAS) networks \cite{TR_38.811} together with new enabling technologies such as large-scale antenna arrays \cite{Rusek13spm}, reconfigurable intelligent surfaces \cite{Renzo20jsac}  and Terahertz communications \cite{Rappaport19ac} constitute  compelling solutions for the emerging services and applications.  Powerful detection schemes play a pivotal role in supporting these  novel techniques for  achieving their potential gains.  Maximum-likelihood (ML) detection is capable of providing the optimal solution by minimizing the probability of error, but it is NP-hard \cite{Verd89alg}, because  its  complexity of finding the exact ML solution grows  exponentially with the size of the constellations and the number of the symbols transmitted, which constitutes a challenge for classical computers. The properties of quantum mechanics such as superposition, entanglement and coherence have been beneficially exploited  in the wireless communication field in terms of quantum information  science \cite{nielsen2002quantum} and quantum teleportation \cite{CacciapuotiTcom20}. Given the intrinsic parallelism of quantum mechanics,  it is promising to investigate the potential of quantum algorithms for solving ML detection problems. 

Specifically, quantum computation  exhibits advantages in solving some problems that require searching through a large space \cite{childs2000finding},  benefiting from the nature of quantum mechanics.  The most canonical examples are  Schor's algorithm  \cite{Shor94}  designed for discrete Logarithms and factoring at an  exponential speedup over classical methods as  well   as   Grover's algorithm \cite{grover1996fast} conceived for unstructured search, which  shows the potential of quadratic speed-up over the classical search. Hence, some  applications of Grover's search inspired quantum algorithms to multi-carrier interleave-division multiple-access (MC-IDMA) systems and to Pareto optimal routing for wireless multihop networks can be found in  \cite{Botsinis13Ac,Alanis18TCOM}. 
 The benefits of these algorithms arise from the assumption of a universal fault-tolerant quantum computer, which is capable of providing  error-free operation. However, a powerful system supporting  millions of physical qubits and high-fidelity long sequences of gate operations is not available at the time of writing, but fortunately,  noisy intermediate-scale quantum (NISQ) devices are likely to become available in the near future \cite{preskill2018quantum}.   The state-of-the-art quantum device size ranges from 50 to 100 qubits. For instance, in 2020 IBM has built a quantum device processing 65 qubits \cite{IBM_roadmap}.
Therefore,  the near-term quantum computers will contain a  limited number of quantum gates due to gate errors and decoherence \cite{Moll2018qst,bharti2021noisy}.  A family of hybrid  quantum-classical algorithms,  namely  variational quantum  eigensolvers (VQE), was developed in \cite{Peruzzo14nature}, which was implemented by combining a reconfigurable quantum device  with a classical computer. The goal of hybrid  quantum-classical algorithms is to take advantage of the potential of NISQ devices by incorporating partial computational resources of  classical computers,  hence it becomes one of leading candidate algorithms for establishing quantum advantages in near-term quantum computers \cite{preskill2018quantum,Benedetti2019qst,cerezo2020variational}.  The quantum approximation optimization algorithm (QAOA) \cite{farhi2014QAOA,farhi2019qaoa}  constitute a hybrid quantum-classical algorithm designed based on the variational principle  for solving combinatorial optimization problems on gate-based quantum computers.  Indeed,  this has become an active field of research due to its promising potential of being implemented by near-term quantum computers  \cite{Hadfield19Alg,farhi2020quantum,Harrigan21NaturePhy} as well as owing to its universality in quantum computation  \cite{lloyd2018quantum,morales2020universality}.  
  %on gate-model quantum computers. 
%Recent advances in the field of quantum computing have spurred 

The basic idea of QAOAs is to alternately apply the problem Hamiltonian, whose ground state encodes the solution of the problem considered, and the initial Hamiltonian.   QAOAs rely on the combination of preparing the parameterized quantum circuit  on the NISQ devices, and a classical optimizer that is used for finding the optimal parameters.  The QAOA was first proposed in  \cite{farhi2014QAOA} for  solving  combinatorial optimization problems, where the paradigm and  analysis on the QAOA were treated in terms of the max-cut problem. As a further advance,  the capability of  the QAOA to solve constrained  combinatorial optimization problems  was studied in  \cite{farhi2015qaoabounded}, where the goal is to maximize the sum of  a series of bounded linear equations. Then,  the performance of the QAOA was analysed in the context of the MAX-$k$XOR and MAX-$k$SAT problems in \cite{lin2016performance}.   The application of the QAOA to the channel decoding of binary codes was presented in \cite{matsumine2019channel} by means of the Ising Hamiltonian  to Boolean constraint satisfaction problems.   Furthermore, inspired by  the adiabatic evolution principle,  a  learning method  was proposed in  \cite{Wecker16PhysRevA} for optimizing parameters aiming at achieving a high overlap to the ground state of the problem Hamiltonian.   Given the high flexibility of QAOAs,  it was extended  in \cite{Jiang17PhysRevA} to solve  Grover's unstructured search problem by replacing Grover's diffusion operator with the transverse field, which only requires single-qubit gates.  In addition, the authors  in  \cite{Hadfield17pems}  proposed a framework for designing the QAOA circuits for solving a couple of combinatorial  problems  subject to  both hard  and soft constraints, such as the graph coloring optimization problem and the traveling salesman problem (TSP).  As a futher development,   the QAOA was generalized as one of the standalone ansatz approaches in \cite{Hadfield19Alg}, termed as the quantum alternating operator ansatz.  
Recall that the quality of the solution produced by QAOAs for a specific  problem of interest  depends on the quality of the variational parameters found by the classical optimizer.  Developing efficient QAOA parameter optimization approaches is therefore of pivotal importance for achieving  quantum advantage. 
Diverse techniques have been conceived  for optimizing the QAOA parameters such as  gradient-based methods \cite{Wang18PhysRevA,Zhou20PhysRevX,crooks2018performance} and gradient-free methods \cite{wecker2016training, Shaydulin19igsc, Streif_2020,Yao20conf}.   
As for the max-cut problem in a general graph,  the analytical expression for a  level-1 QAOA  was provided for guiding  the associated parameter selections in \cite{Wang18PhysRevA}. Furthermore,  analytical expressions were also  derived for the QAOA haivng an arbitrary number of levels for a class of special max-cut graph instances  in \cite{Wang18PhysRevA}.  Moreover, a number of methods based on neural networks and reinforcement learning have been applied for optimizing the QAOA parameters  \cite{crooks2018performance, wecker2016training, Shaydulin19igsc, Streif_2020, Yao20conf}, concerning different max-cut graph instances.

Given the high flexibility of QAOAs in terms of handling various coherence times and gate requirements etc, there has been a growing interest in exploring the advantages of QAOAs in solving numerous  practical problems of diverse fields. 
In this paper,  we discuss how  to apply the QAOA for solving  ML detection problems that are NP-hard in classical computers.   
The basic procedure of the QAOA is to alternately apply the problem Hamiltonian and the initial Hamiltonian -- also known as mixing  Hamiltonian \cite{Hadfield19Alg} or driver Hamiltonian \cite{Hen16PhysRevA} -- to the initial state of the quantum system.   The fundamental principle of QAOAs relys on gradually driving the quantum system to its ground state  based on the quantum adiabatic evolution technique of \cite{farhi2000quantum}.    
Explicitly, the evolution of a quantum system is governed by  the Schr\"{o}dinger equation, and the adiabatic theorem tells us how to track this evolution, when the system changes sufficiently slowly.  To  elaborate a little further,   the quantum adiabatic algorithm  starts from an initial Hamiltonian whose ground state is easy to prepare, and evolves to a final Hamiltonian whose ground state encodes the solution of the problem considered.  For implementing the QAOA,  the   evolution is encoded into a series of unitary quantum logic gates. Therefore, the goal of this paper is to   illustrate the applicability of QAOAs to the family of ML detection problems and hence to open up new avenues of optimizations in wireless communications.  We commence  with  a brief overview  of the  applications  of  QAOAs in solving different  problem instances, as seen  in {\bf Table \ref{Tab:Comspaper}}.  In this paper we present the first results on the QAOA  applied to a real-world  ML detection problem in wireless communications.  Our main contributions are summarized as follows.
 
\begin{enumerate}
\item   We encode the ML detection problem of binary symbols transmitted over a multiple-input and multiple-output (MIMO) channel into a Hamiltonian operator in the Ising transverse field  \cite{sachdev2007quantum}. Specifically, we use a qubit to encode a single binary symbol. Hence the total number of  qubits to be processed by the QAOA is equal to the number of parallel   data symbols transmitted. Correspondingly, any legitimate solution of the ML detection problem can be represented by  a sequence of qubits and the optimal solution is encoded into  the ground state of the problem Hamiltonian.

\item  For  illustrating the fundamental  principles of the QAOA, we provide both analytical and numerical results concerning the evolution of the eigenvalues in terms of the ML detection problem using a time-dependent Hamiltonian, which is a linear interpolation between the initial Hamiltonian and the problem Hamiltonian. 

\item We transform the ML detection problem into  a $p$-level QAOA circuit, where  in each level the problem Hamiltonian that encodes the objective function of the ML detection problem is applied  to the initial state of the quantum system followed by  the initial Hamiltonian.  The $2p$ times Hamiltonian operators involve the variational  parameters, which are optimized classically for best performance. In particular, 
we derive the analytical expression of the ML detection problem for the level-1 QAOA, which simplifies the numerical optimization for the optimal values of the parameters.

\item We provide  numerical  results for characterizing the expectation values of the  QAOA solution of the ML detection problem, which illustrates that the energy landscape of the QAOA is nonconvex and has locally optimal points.   Furthermore,  the performance of the QAOA based ML detector  is evaluated compared to that of both the classical ML  detector as well as to the classical minimum mean squared error (MMSE) detector.
\end{enumerate}

\begin{table*}[!htb]
\renewcommand{\arraystretch}{1.5}
\centering
\caption{Problem instances solved by QAOAs}
\label{Tab:Comspaper}
\begin{center}
 {\small
\resizebox{\textwidth}{!}{
\begin{tabular}{| l | c | c | c | c | c | c | c | c | c | c | c | c |}
    \hline
    {} & \cite{farhi2014QAOA} -2014 &  \cite{farhi2015qaoabounded}-2015  & \cite{lin2016performance}-2016   &\cite{Jiang17PhysRevA}-2017 & \cite{Hadfield17pems}-2017 & \cite{Wang18PhysRevA}-2018 &\cite{crooks2018performance}-2018 & \cite{Shaydulin19igsc}-2019 & \cite{Hadfield19Alg}-2019 & \cite{bravyi2020hybrid}-2020 & \cite{Harrigan21NaturePhy}-2021& This work\\
   \hline
	Max-Cut          &  {$\surd$} & {$\surd$} &  {$\surd$ } & {  } & {$\surd$} & {$\surd$}   & { $\surd$ } & {$\surd$}  & { $\surd$ }&  { } & {$\surd$ }& { } \\ \hline
	 MIS    &  {       } & {    } & {       } & { } & {} & {}   & {      } & {    } & {$\surd$}& {  } &{ }&{ } \\ \hline
	Graph Coloring   &  {       } & {       } & {       } & {        } & {$\surd$ } & {       }   & {       } & {   } &{$\surd$} & { $\surd$ }  &{ } & { } \\ \hline
	TSP             &  {} & {       } & {} & {        } & {       } & {       }   & {  } & {    }&{$\surd$} &{  } &{} &{}\\ \hline
%	2/3-SAT     &  {       } & {       } & {       } & {        } & {       } & {       }   & {        } & {    } &{      } & {   }\\ \hline
Unstructured search	  &  {       } & {       } & {       } & { $\surd$ } & {       } & {       }   & {        } & {    } &{      } & {   }&{}&{}\\ \hline
ML detection 	  &  {       } & {       } & {       } & {  } & {       } & {       }   & {        } & {    } &{      } & {   }&{} &{$\surd$}\\ \hline
	\end{tabular}}}
\end{center}
\end{table*}

\subsubsection*{Organization}

The rest of our paper is organized as follows. Section \ref{sec:sysmod} presents the problem model of the ML detection, followed by modelling its quantum Hamiltonians in Section \ref{sec:ham_Adia}.  Section \ref{sec:qae} introduce the quantum adiabatic evolution for characterizing the ML detection based quantum system.  
Then, Section \ref{sec:qaoa}  discuss the procedure of the QAOA for solving the ML detection problem.  The computational complexity analysis of the QAOA for solving the ML detection problem is provided in Section \ref{sec:comanal}.
Finally, Section \ref{sec:sim} presents  our simulation results and discussions, followed by our conclusions in Section \ref{sec:conclu}.

\section{System Model}
\label{sec:sysmod}

Consider a MIMO system and transmitting binary symbols  over an ${M_r \times M_t}$  channel matrix,  having $M_t$ transmit antennas (TA) as well as  $M_r$ receive antennas (RA).  Let  $\mathbf{H} \in  \mathcal{R}^{M_r \times M_t}$ be the channel matrix from the transmitter to the receiver, which is assumed to be known to the receiver.  Therefore, the received signal  can be expressed as
\begin{equation}
\mathbf{y} = \mathbf{H}\mathbf{s} + \mathbf{n},
\end{equation}
where $\mathbf{y} \in \mathcal{R}^{M_r}$   and $\mathbf{s}  \in \mathcal{R}^{M_t}$ are the vectors of  the received  and the transmitted  signals, respectively, while    $\mathbf{n} \sim \mathcal{N}(0, \mathbf{I}_{M_r})$ denotes the noise vector. 

The maximum likelihood (ML) detector can be formulated as 
\begin{eqnarray}
\begin{aligned}
\min_{\mathbf{s} \in \mathcal{X}^{M_t}} \| \mathbf{y} - \mathbf{H}\mathbf{s}\|^2,  \label{eq:ml_1}
\end{aligned}
\end{eqnarray}
where $\mathcal{X}$ represents the signal constellation of the binary symbols.  Explicitly,  problem \eqref{eq:ml_1}   can be physically interpreted  as  follows:  The ML detector  is to find a vector $\mathbf{s}$ producing a vector $\mathbf{H}\mathbf{s}$,  which is closest to the received signal $\mathbf{y}$. Correspondingly,  problem \eqref{eq:ml_1} can be reformulated as the following optimization form:
\begin{eqnarray}
\begin{aligned}
 \min_{\mathbf{s} \in \mathcal{X}^{M_t}} \mathbf{s}^{T}\mathbf{H}^{H} \mathbf{H}\mathbf{s} - 2\mathbf{y}^{T} \mathbf{H}\mathbf{s}  + \mathbf{y}^T \mathbf{y}, \label{eq:ml_2} 
\end{aligned}
\end{eqnarray}
where the superscript $T$ denotes the  transpose of a matrix. Note that problem \eqref{eq:ml_2} is NP-hard due to the discrete constraints $\mathcal{X}^{M_t}$ \cite{Verd89alg}.

Therefore, the objective function of  problem  \eqref{eq:ml_2} can be expressed for BPSK having $\mathcal{X} = \{-1, +1\}$  as 
\begin{eqnarray}
\begin{aligned}
f(\mathbf{s}) &=   \mathbf{s}^{T}\mathbf{H}^{T} \mathbf{H}\mathbf{s} - 2 \mathbf{y}^{T} \mathbf{H}\mathbf{s}  + \mathbf{y}^T \mathbf{y}   \\ 
& = \sum_{k,l = 1}^{N} A_{k,l}s_k s_l  - \sum_{k=1}^{N} 2 b_k s_k + c \\
 &= \sum_{l>k}^{N} 2 A_{k,l} s_k s_l  - \sum_{k=1}^{N} 2 b_k s_k + c + \sum_{k=1}^N A_{k,l}, \label{eq:f_bpsk}
\end{aligned}
\end{eqnarray}
where $N = M_t$, $\mathbf{A} = \mathbf{H}^{T} \mathbf{H}$, $\mathbf{b} = \mathbf{y}^{T}\mathbf{H}$ and $c = \mathbf{y}^T \mathbf{y}$.    Consequently, $A_{k,l}$ and $b_k$ are the $(k,l)$-th element and the $k$-th element in $\mathbf{A}$ and $\mathbf{b}$, respectively.  The last step in \eqref{eq:f_bpsk} comes from the fact that $A$ is Hermitian, i.e.,  $A_{k,l} = A_{l,k}$, $\forall k, l$.

\section{ Quantum Hamiltonians and the Adiabatic Theorem}
\label{sec:ham_Adia}

\subsection{Hamiltonian $H_f$ of the ML Problem}
\label{subsec:Hf}
\begin{figure} [htp!]
\centering
\includegraphics[clip,width = 0.8\textwidth]{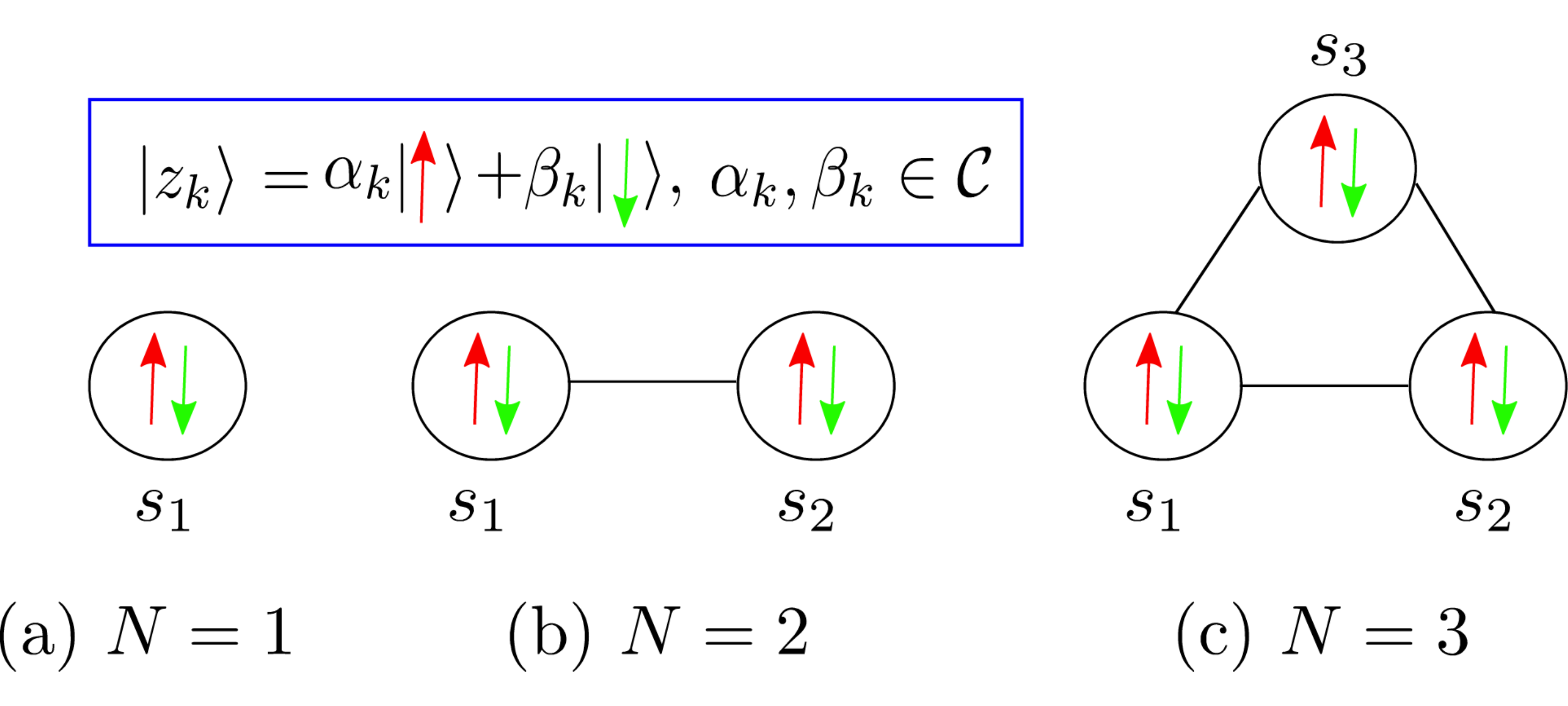}
  \caption{An example of the connections between different symbols received, where ${\color{red}\uparrow}$ and ${\color{green}\downarrow}$ represent the two possibilities for BPSK systems, respectively. Due to the quantum superposition, an unknown symbol $s_k$ can be represented as a single qubit $|z_k\rangle = \alpha_k |{\color{red}\uparrow} \rangle + \beta_k| {\color{green}\downarrow} \rangle, \alpha_k,\beta_k \in \mathcal{C}$. }
 \label{fig:ex_3qubit}
   \vspace{-1.0em}
\end{figure}
Since binary symbols are considered,  we can use a single qubit to encode the values of a  single symbol transmitted. Hence the total number of  qubits required for quantum computation is equal to the number of parallel  data symbols transmitted. More explicitly, an example is provided in Fig.  \ref{fig:ex_3qubit} for demonstrating the connections between the different symbols received, where ${\color{red}\uparrow}$ and ${\color{green}\downarrow}$ represent the two legitimate states of BPSK systems, respectively. However, in quantum computing one qubit can be represented by  a superposition of the two states.  Hence, given an unknown binary  symbol $s_k$,  $k = 1,\cdots,N$, the associated quantum state $|z_k\rangle$ can be expressed as    $|z_k\rangle = \alpha_k |{\color{red}\uparrow} \rangle + \beta_k| {\color{green}\downarrow} \rangle, \alpha_k,\beta_k \in \mathcal{C}$. 

In order to transform the ML detection problem from a classical computation to   quantum computation,  we map \eqref{eq:f_bpsk} to its spin Hamiltonian.   Explicitly, to arrive at the Hamiltonian of \eqref{eq:f_bpsk}, we define $z_k \in \{0,1\} $ as a spin-$\frac{1}{2}$ qubit associated with 
\begin{eqnarray}
\begin{aligned} \label{eq:z_basis}
|0\rangle = \begin{pmatrix}
1 \\0
\end{pmatrix}, ~\text{and}~~
|1\rangle = \begin{pmatrix}
0 \\1
\end{pmatrix}.
\end{aligned}
\end{eqnarray} 
Then, we have 
\begin{eqnarray}
\label{eq:pauli_z}
\begin{aligned}
\sigma_z^{(k)} |z_k\rangle = \pm 1|z_k\rangle, ~\text{and}~~
\sigma_z^{(k)} = \begin{pmatrix}
1&0 \\
0&-1
\end{pmatrix}.
\end{aligned}
\end{eqnarray}
Note that  $\sigma_z^{(k)}$ denotes  the Pauli-Z operator acting on the $k$-th qubit.   We can see that the binary symbol $s_k$ is mapped  onto the eigenvalues of the Pauli-Z operator.
Then, we have the Ising Hamiltonian associated with  \eqref{eq:f_bpsk} as follows.
\begin{eqnarray}
\begin{aligned} \label{eq:Hp_f}
H_f  =  \sum_{l>k}^N 2A_{k,l} \sigma_z^{(k)}  \sigma_z^{(l)} -  \sum_{k=1}^{N}  2 b_{k} \sigma_z^{(k)}  + c + \sum_{k=1}^N A_{k,k}.
\end{aligned}
\end{eqnarray}
Note that the problem Hamiltonian $H_f$ was also referred to as phase Hamiltonian in  \cite{Hadfield19Alg}. Now we verify that the problem Hamiltonian $H_f$ such that the  eigenvectors of $H_f$ forms the solution space of the original problem \eqref{eq:ml_1}. 

\begin{proposition}\label{pro:verfy_Hp}
 Let $|z\rangle = |z_1,\cdots,z_N \rangle$ be the eigenvector of  the problem Hamiltonian $H_f$, where $\{|z_1,z_2, \cdots,z_N \rangle$, $z_k \in\{0,1\} \}$ is a basis for the $2^N$-dimensional Hilbert space of the quantum computer.  The solutions $\{|s_1,\cdots, s_N \rangle,s_k \in \{-1,+1\} \}$ of the ML  detection  problem \eqref{eq:ml_1} can be mapped to the eigenvectors of $H_f$ by a bijective function such that $g:(\{|z_1,z_2, \cdots,z_N \rangle \} \rightarrow (\{|s_1,s_2, \cdots,s_N \rangle \}$. 

\begin{proof}
See Appendix~A.
\end{proof}
\end{proposition}

Having encoded the MIMO-ML detection problem into its Hamiltonian, we can now expand it to a more general multi-user systems. 
\begin{proposition} \label{pro:mu_mimo}
The encoding process of the  problem Hamiltonian to the MIMO-ML detector  is suitable for  the multi-user single-input and single-output (SISO)/MIMO systems of binary symbols.
\begin{proof}
See Appendix~B
\end{proof}
\end{proposition}

\subsection{The Adiabatic Theorem}
The adiabatic theorem  was first proved in 1928 \cite{Born1928Phy} for describing certain properties of particle behaviour in quantum systems, which can be formulated as follows. 
\begin{theorem}[\cite{Scott_lec19}]
 Consider a time-varying Hamiltonian $H(t)$, which starts from $H_I = H(0)$ at $t=0$ and subsequently becomes $H_{t'}$ at some later time $t = t'$.  If a quantum system is initially in the ground state $H_I$ and as long as the change in the Hamiltonian is sufficiently slow, the system state is likely to remain in the ground state throughout the evolution, therefore being in the ground state of $H_{t'}$ at $t = t'$.
\end{theorem}
The adiabatic theorem tells us that  the quantum system can smoothly evolve from a known ground state of Hamiltonian $H(0)$ to an unknown ground state of Hamiltonian $H(T)$ given a run time $T$, where the state of the system  evolves according to the Schr\"{o}dinger equation \cite{griffiths2018introduction}:
\begin{align}
\label{eq:TDSE}
\frac{id |\psi(t)\rangle }{dt} = H(t) |\psi(t)\rangle.
\end{align}
Here the state $|\psi(0)\rangle$ at $t=0$ is known and easy to construct.  The Hamiltonian that governs the evolution is given by 
\begin{eqnarray}
H(t) = \tilde{H}(\frac{t}{T}),
\end{eqnarray}
where $T$ controls the variation rate of $H(t)$.  Note that a Hamiltonian is an operator described by a Hermitian matrix, whose eigenstates and the associated eigenvalues represent the states of the system and the corresponding energy levels, respectively.   Based on the quantum adiabatic  theorem, the state of the quantum computer $| \psi(t) \rangle$ is  close to the ground state of $H(t)$ for $0\le t\le T$, and in particular $|\psi(T)\rangle$ will be close to the ground state of $H_f = H(T)$, i.e.  the encoded solution of the problem considered.  Consequently, following the  quantum adiabatic  theorem, the evolution of a system can be treated as a time independent Schr\"{o}dinger equation \cite{griffiths2018introduction}, i.e., 
$H|\psi\rangle = \lambda |\psi \rangle$.

\subsection{Initial Hamiltonian $H_B$ }
We now consider a $N$-qubit Hamiltonian $H_B$ whose ground state is easy to find.  Define the states $|x_k\rangle$ as the eigenstates of the x-component of the $k$-th spin-$\frac{1}{2}$, where  
\begin{eqnarray}
\begin{aligned}
|x_k=0\rangle = \frac{1}{\sqrt{2}}\begin{pmatrix}
1\\1
\end{pmatrix}, ~~\text{and}
|x_k=1\rangle = \frac{1}{\sqrt{2}}\begin{pmatrix}
1\\-1
\end{pmatrix}
\end{aligned}.
\end{eqnarray}
Consequently, we have 
\begin{eqnarray}
\begin{aligned}
\sigma_x^{(k)}|x_k  \rangle = x |x_k\rangle, ~\text{and}~~ \sigma_x^{(k)} = \begin{pmatrix}
0 & 1 \\ 1 & 0
\end{pmatrix},
\end{aligned}
\end{eqnarray}
where $x = +1, -1$ and $\sigma_x^{(k)}$ is the Pauli-X operator acting on the $k$-th qubit. Therefore, we can express the initial Hamiltonian of \eqref{eq:f_bpsk}  \cite{farhi2014QAOA,Hadfield19Alg} as 
\begin{eqnarray}
\label{eq:H_B}
H_B = \sum_{k=1}^{N} \sigma_x^{(k)}.
\end{eqnarray}
The ground state of $H_B$ is $|x_0 = 0\rangle |x_1 = 0\rangle \cdots |x_{N} = 0\rangle$. This state can be rewritten as an equiprobable superposition of  the Z-basis states. Hence, the initial state is given by 
\begin{eqnarray}
\begin{aligned}
|\psi(0)\rangle =& |x_1 = 0\rangle |x_1 = 0\rangle \cdots |x_{N} = 0\rangle \\
=& \frac{1}{\sqrt{2^N}}\sum_{z_1}\cdots \sum_{z_{N}} |z_1\rangle \cdots |z_{N}\rangle,  \label{eq:psi0}
\end{aligned}
\end{eqnarray}
where $z_k \in \{0,1\}$ and  $|x_k = 0\rangle = \frac{1}{\sqrt{2}} (|0\rangle + |1\rangle)$.

\section{Quantum Adiabatic Evolution}
\label{sec:qae}
\subsection{The Adiabatic Evolution}
\label{sec:sub_ae}
Following the adiabatic evolution, we assume that the quantum system starts from a known ground state $H_B$ and evolves to  an unknown ground state $H_f$. Consider  a linear interpolation between $H_B$ and $H_f$, formulated as 
\begin{eqnarray}
H(t) = (1 - \frac{t}{T}) H_B  + H(\frac{t}{T})H_f.
\end{eqnarray}

Let $\tau = \frac{t}{T}$,  $0\le \tau \le 1$. We then have the single parameter Hamiltonian evolution of 
\begin{eqnarray}
\tilde{H}(\tau) = (1 - \tau) H_B  + \tau H_f.
\end{eqnarray}

Let us prepare the system by ensuring that  it evolves at $t=0$ from the ground state of $H(0) = H_B$. According to the adiabatic theorem, if $g_{min}$ is not zero and the system evolves following \eqref{eq:TDSE}, then for  a sufficiently long time $T$,  $|\psi(T) \rangle$ will be close to the ground state of $H_f$, which is the encoded solution of the  problem considered.

Upon defining the instantaneous eigenstates and the associated eigenvalues of $\tilde{H}(\tau)$ by
\begin{eqnarray}
\tilde{H}(\tau) |\psi_l(\tau)\rangle = \lambda |\psi_l(\tau)\rangle,
\end{eqnarray}
where $|\psi_l(\tau)\rangle$ represents the $l$-th eigenstates of $\tilde{H}(\tau)$ and  $\lambda_l(\tau)$ is the associated eigenvalue.  Furthermore,  $\lambda_l(\tau)$ represents the  energy levels of  the quantum system, which should be sufficiently well separated, satisfying 
\begin{equation}
\lambda_0(\tau) \le \lambda_1(\tau) \le \cdots \le \lambda_{N-1}(\tau) .
\end{equation}

According to the adiabatic theorem, if $\lambda_1(\tau) - \lambda_0(\tau)> 0$,  then  the expectation value  obeys 
\begin{eqnarray}
\lim_{T\rightarrow \infty} \langle \psi_0 (1)| \psi(T) \rangle = 1,
\end{eqnarray}  
which means that there exists a $\psi(t)$ obeying \eqref{eq:TDSE} that is very close to the instantaneous ground state of $H(t)$ with a non-zero gap, if $T$ is long enough. 

Upon considering a particular search problem, the quantum algorithm is considered to be successful if the  run time required only at most increases polynomially with the number of bits. 
As discussed in \cite{farhi2000quantum}, the  run time  required is related to the spectrum of $\tilde{H}(\tau)$,  which has to satisfy that 
\begin{equation}
T \gg \frac{\xi}{g^2}, \label{eq:T}
\end{equation}
where $g$ is the minimum gap between the two lowest eigenvalues $\lambda_1(\tau)$ and $\lambda_0(\tau)$, which is formulated as
\begin{eqnarray}
g = \min_{0\le \tau \le 1} \lambda_1(\tau) - \lambda_0(\tau). \label{eq:min_gap}
\end{eqnarray}
Furthermore,  
$\xi$ in \eqref{eq:T} is no higher than the largest eigenvalue of $H_f - \tilde{H}(0)$, which is usually a polynomially increasing function of $n$,  and thus $T$ is dominated by $g^{-2}$ \cite{farhi2000quantum,mcgeoch2014adiabatic}.

\subsection{Single-qubit Example}
\label{sec:sub_one_ex}
Consider a single-qubit problem associated with $M_t = M_r = N=1$. Then,  \eqref{eq:f_bpsk} can be rewritten as 
\begin{eqnarray}
\begin{aligned} \label{eq:one_qb1}
f(s) = a s^2 -2 b s + c,
\end{aligned}
\end{eqnarray}
where $a =   |h|^2, b = yh$ and $c =  |y|^2$. As $s \in \{-1,+1\}$, $s^2 = 1$. Thus, \eqref{eq:one_qb1} can be cast as 
\begin{align} \label{eq:one_qb2}
f(s) = -2b s + a + c.
\end{align}

Consequently, the  Hamiltonian $H_f$ of problem \eqref{eq:one_qb2} can be  expressed as 
\begin{eqnarray}
\begin{aligned}
H_f = -2b \sigma_z + a+c = \mathrm{diag}([a-2b+c, a+2b+c]). \label{eq:Hp_oneqbit}
\end{aligned}
\end{eqnarray}
 We can see that $H_f$ has two different eigenvalues, namely $\lambda = a-2b+c$ associated with the eigenstate $|0\rangle$ and $\lambda' = a-2b+c$  associated with the eigenstate $|1\rangle$, respectively. This means that the ground state $\psi(\tau=1)$ of $H_f$ can be expressed as 
 \begin{eqnarray}
 \begin{aligned}
 \psi_0(\tau =1) = \begin{cases}
 |0\rangle & \text{if}~ \lambda \le \lambda'  \\
 |1 \rangle &  \text{if}~ \lambda >  \lambda' 
  \end{cases}.
 \end{aligned}
 \end{eqnarray}
Furthermore, from \eqref{eq:H_B}, we have $H_B = \sigma_x$.  Thus, the smooth  interpolating Hamiltonian of 
 \begin{eqnarray}
 \begin{aligned}
 \tilde{H}(\tau) &= \begin{pmatrix}
 (a-2b+c)\tau& 1-\tau \\ 
 1- \tau           &  (a+2b+c)\tau
 \end{pmatrix}
 \end{aligned}
 \end{eqnarray}
has two eigenvalues $(a+c)\tau \pm \sqrt{1-2\tau + (1+4b^2)\tau^2}$, which will be further discussed  in Section \ref{sec:sim}.

\subsection{Two- and Three-qubit Example}
\label{sec:sub_tt_ex}
Let us first consider a two-qubit example associated with a MIMO system of $N=2$,  which allows the signal values $\{-1,-1\}$, $\{-1,+1\}$, $\{+1,-1\}$ and $\{+1,+1\}$.   
Correspondingly, from  \eqref{eq:Hp_f}, we have  $H_f$ for $N = 2$ as follows
\begin{dgroup*}
\begin{dmath} \label{eq:Hp_f_twoqbit}
H_f = 2A_{1,2}  \sigma_z^{(1)}\sigma_z^{(2)} - 2(b_1 \sigma_z^{(1)}+b_2\sigma_z^{(2)} ) +   c +  A_{1,1} + A_{2,2}.
\end{dmath}
\end{dgroup*}
Furthermore, we take $H_B$  from  \eqref{eq:H_B} in conjunction with $N=2$,  yielding 
\begin{eqnarray}
\begin{aligned}
\label{eq:H_B_2qubit}
H_B &= \sigma_x^{(1)}+\sigma_x^{(2)} ,
\end{aligned}
\end{eqnarray}
which has the minimum eigenvalue of -1 associated with the eigenstate $|x_1 = 0\rangle|x_2 = 0\rangle$.  The full matrix forms of $H_f$, $H_B$ and $\tilde{H}(\tau)$ are given in \emph{Appendix~C}.  
Therefore,  the corresponding interpolating Hamiltonian, namely $\tilde{H}(\tau) = (1 - \tau)H_B + H_f$, can be written as the sum of  terms in \eqref{eq:Hp_f} and \eqref{eq:H_B}, where each term acts on two qubits.

Finally, we consider a three-qubit system, i.e. $N = 3$. The corresponding interpolated  Hamiltonian, $\tilde{H}(\tau)$ can be written as the sum of 
\begin{eqnarray}
\begin{aligned}
H_f &=  2\sum_{(k,l) \in E} \sigma_z^{(k)}\sigma_z^{(l)} - 2\sum_{k}\sigma_z^{(k)} +  \tilde{c}\\
H_B &= \sum_{k=1}^3\sigma_x^{(k)},
\end{aligned}
\end{eqnarray}
where $E = \{(1,2),(1,3),(2,3)\}$ and $\tilde{c} = c+ \sum_{k=1}^3  A_{k,k}$.  The eigenvalues of $\tilde{H}(\tau)$ for  two-qubit and three-qubit ML problems will be further discussed in Section \ref{sec:sim}.

\section{QAOA for Solving the ML Detection Problem}
\label{sec:qaoa}
In this section, we first present the basic principles of quantum adiabatic approximation using Trotterization\footnote{ Trotterzation is a very useful tool for simulating non-commuting operators in quantum computers, by using the Trotter-Suzuki formula \cite{trotter1959product,suzuki1976relationship} to approximately decompose the system operator into a sum of  easy to implement operators.} in  quantum computers \cite{Wu02PhysRevLett}, which usually involves a long sequence of gates.  For avoiding this issue,  we  discuss how the QAOA may be adapted for solving the ML detection problem. 

\subsection{The Quantum Adiabatic Approximation}
Recall that the evolution of a quantum system  is governed by the Schr\"{o}dinger's equation \eqref{eq:TDSE}.  If the system starts from  some initial  state $|\psi(0)\rangle$,  the solution to  \eqref{eq:TDSE}  is the unitary evolution of the  state \cite{griffiths2018introduction}, which is given by
\begin{eqnarray}
\begin{aligned}
|\psi(t)\rangle = e^{-i H t} |\psi(0)\rangle = e^{-i(H_f  + H_B) t}|\psi(0)\rangle, \label{eq:sol_tdse}
\end{aligned}
\end{eqnarray}
 describing what state the  quantum system will be in after $H$ has been applied to it over a certain time period of $t$.   As a result,  from  a classical computing perspective,  the adiabatic algorithm  is the process where  the state $|\psi(T) \rangle$ is obtained by applying a series of unitary operators to the initial state. In the adiabatic evolution algorithm, the  unitary operator $U(H,T)$ is  approximated by a product of unitary operators relying on a discrete-time basis by discretizing the interval $[0,T]$ into $p$  slices denoted by  $\mathcal{T} = \{t_1, \cdots,t_p\}$. Then, Trotterization is applied to approximate each discrete time slice. Specifically, the basic idea of  the Trotterization technique is to decompose the system Hamiltonian into a sequence of short-time operators that are easy to simulate, and then approximate the total evolution by consecutively simulating each simpler operator.  Next we first introduce the method of implementing $U(H,T) = e^{-i H t}$ at the discrete time instant $t$, $t \in  \mathcal{T}$. 

\begin{remark}
The operators $H_f$ and $H_B$ do not commute, i.e.  $[H_f,H_B] = H_f H_B - H_B H_f \neq 0$.
\end{remark}
The commutator of  $\sigma_z$ and $\sigma_x$ can be calculated as $[\sigma_z,\sigma_x] = -[\sigma_x,\sigma_z] = 2i\sigma_y$, hence we see that $\sigma_x$ and $\sigma_z$ do not commute. Since $H_f$ and  $H_B$ are functions of $\sigma_z$ and $\sigma_x$, respectively,  we have  $H_f H_B - H_B H_f \ne 0$, i.e.  $H_B$ and $H_f$ do not commute. As a consequence,  the following matrix exponentials have to obey  $e^{-i(H_f  + H_B) t} \neq e^{-i H_f t}  e^{-i H_B t}$. Based on Trotter product formula \cite{Chin02JSP},  we have 
 \begin{eqnarray}
 \begin{aligned}
 e^{-i H t} = e^{-i(H_f  + H_B) t}  = \lim_{ r \rightarrow \infty} \left( e^{-i H_f t / r}  e^{-i H_B  t / r} \right)^r, r \in \mathcal{Z}.  \label{eq:trotter_formula}
 \end{aligned}
 \end{eqnarray}
Therefore, the unitary operator $U(H,t)=  e^{-i H t} $ can be approximated by  a sequence of small slices. Since the bound of \eqref{eq:trotter_formula} is only approached  at $r \approx \infty$, we  have to truncate the series at a  finite order, such as a finite number $r$, for simulations on quantum computers.  For each slice $e^{-i H_f t / r}$, there is an approximation based on the Trotter-Suzuki formula \cite{trotter1959product,suzuki1976relationship}, which is formulated as 
\begin{eqnarray}
\begin{aligned}
 e^{-i(H_f  + H_B) t /r} = e^{-i H_f t / r}  e^{-i H_B  t / r} + \epsilon,  \label{eq:trotter}
\end{aligned}
\end{eqnarray}
where $\epsilon$ is the approximation error.  From the Baker–Campbell–Hausdorff  formula of \cite{nielsen2002quantum},  the norm of the error obeys $\| \epsilon \|_2 \le O( \| i H_f t / r \| \cdot \| i H_B  t / r \| ) = O\big( \frac{t^2}{r^2}\| H_f \| \|  H_B \| \big)$.  

 Therefore, the unitary operator $U(H,T) = e^{-i H T}$ can be written as a product of $p$ unitary operators that are easy to simulate, 
 \begin{eqnarray}
 \begin{aligned}
 U(H,T) &= e^{-iHT} \\&= U(H,t_p =  T - \bigtriangleup)  U(H,t_{p-1} = T - 2\bigtriangleup)  \cdots   U(H,t_1 = 0) \\
 & = \prod_{k = 1}^p  \big(e^{-i H_f t_k / r}  e^{-i H_B  t_k / r}\big)^r,
 \end{aligned}
 \end{eqnarray} 
where we have $U(H,t_k) = e^{-i H_f t_k / r}  e^{-i H_B  t_k / r}$ and $\bigtriangleup = \frac{T}{p}$.  We  can see that implementing these unitary operators requires  a quantum circuit of  depth  $2pr$, which indicates that the length of the circuits depends both on the evolution time and on the Trotter steps. As a result,  the depth of the associated quantum circuits grows with the product of $p$ and $r$,   which usually requires a very long circuit. Gate errors and decoherence restrict the number of sequential gate operations in the  quantum devices,  therefore a class of  hybrid classical quantum algorithms are developed for near-term quantum computing \cite{Moll2018qst,bharti2021noisy}.  Indeed, QAOA belongs to the family of  hybrid classical quantum algorithms,  which creates a parameterized quantum state by alternately applying the Hamiltonian $H_f$ and $H_B$  $p$ times for a  given $p$.

\subsection{Implementation of QAOA}
\label{sub:imqaoa}
\begin{figure} [!t]
\centering
\subfigure[Optimizations of $F_p$ based on VQE principles.]{
\includegraphics[clip, width = 0.5\textwidth]{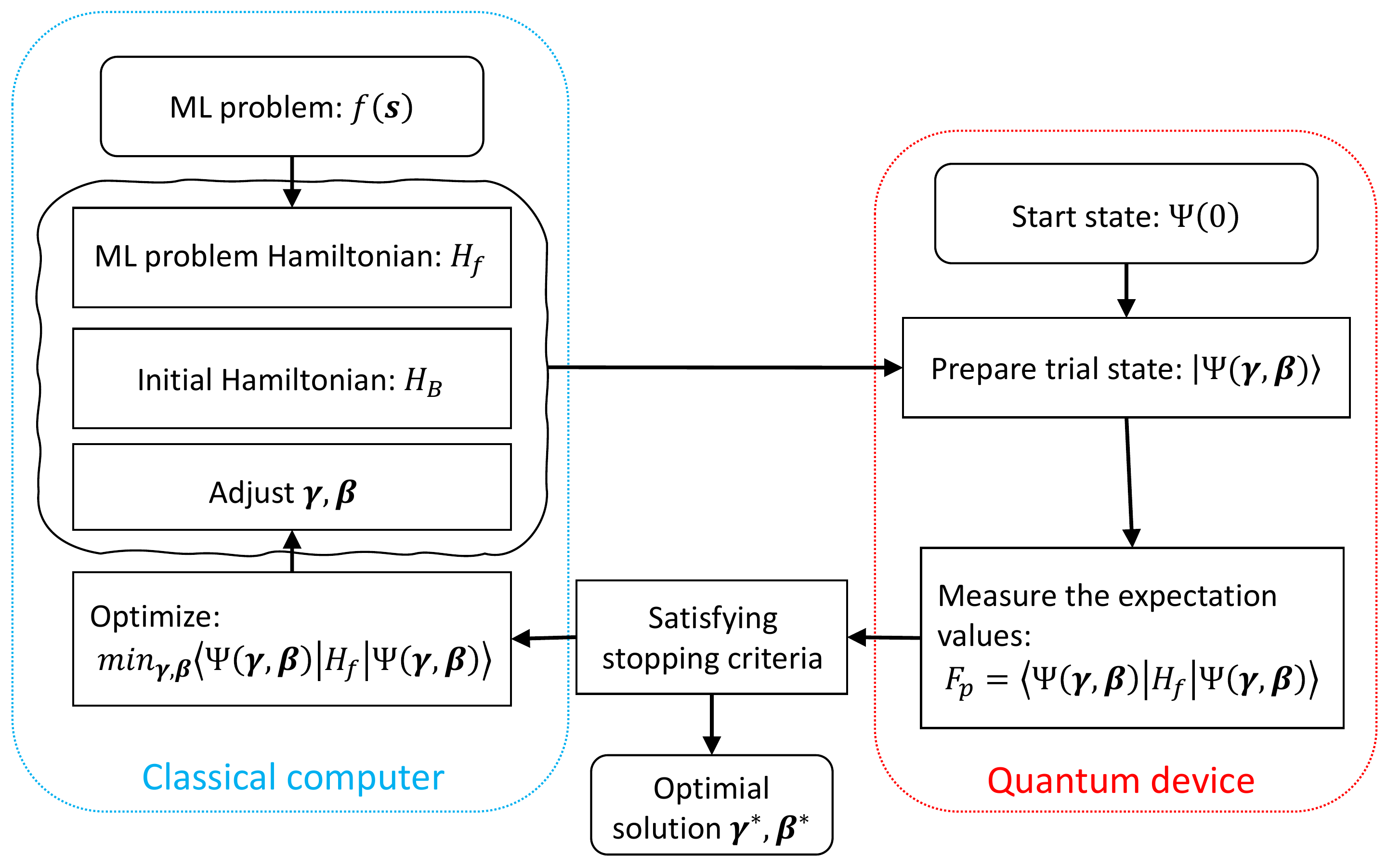}
%\vspace*{-0.5em}   \caption{ QAOA diagram. The realization of QAOA compose of both a quantum processor and a classical processor.  }
 \label{fig:qaoa_diag_vqe}
}
\subfigure[Optimizations of $F_p$ using classical solver only. ]{
\includegraphics[clip, width = 0.38\textwidth]{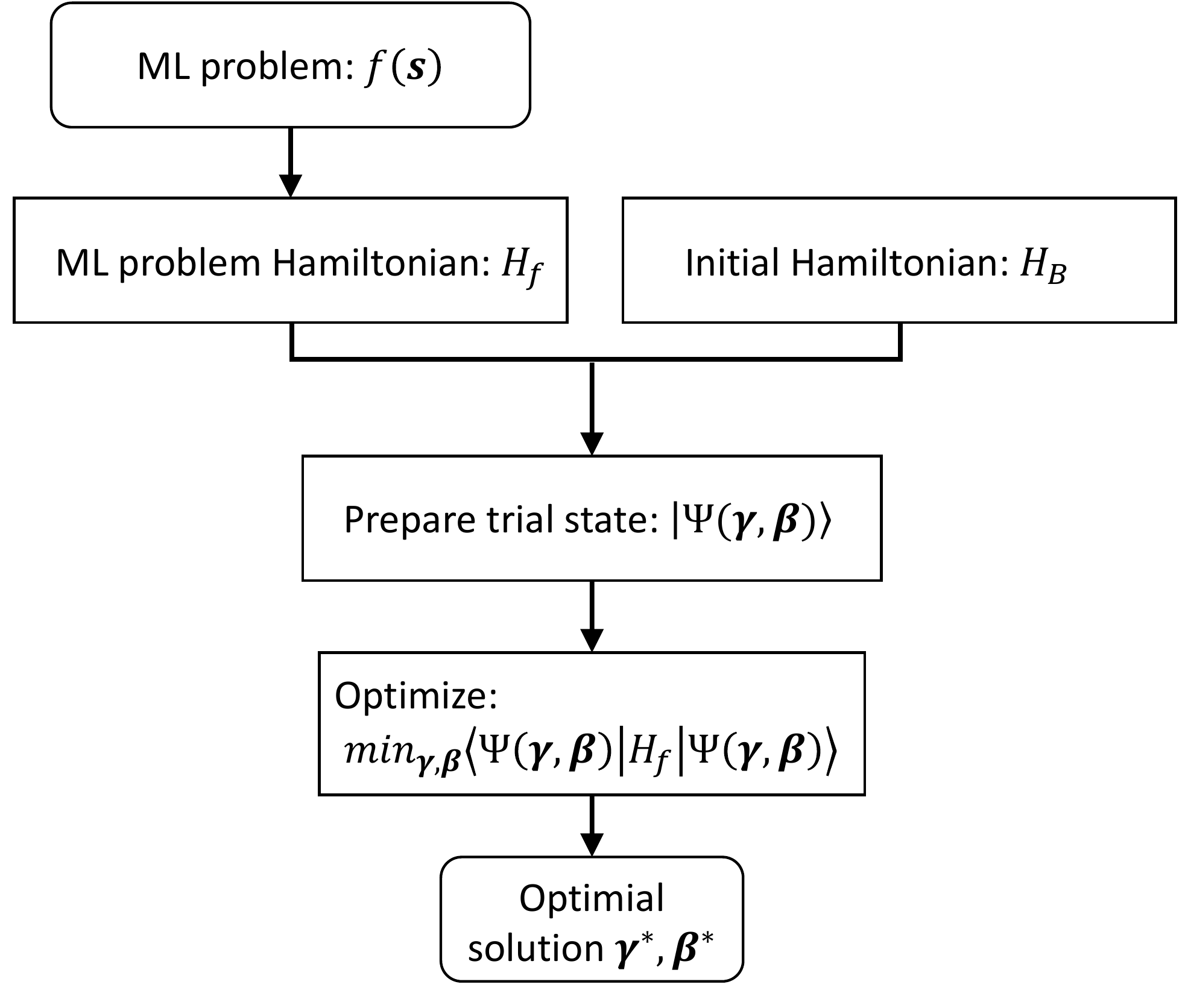}
\label{fig:qaoa_diag_clas}
}

 \caption{QAOA diagram for optimization $F_p$.}
\label{fig:qaoa_diag}
   \vspace{-1.5em}
\end{figure}
Instead of using  Trotterization methods \cite{Wu02PhysRevLett,Sun_2020iop},  QAOA  prepares a pair of unitary operators in terms of $H_f$ and $H_B$.   For ease of computation, we introduce the following remark to simplify $H_f$.
 
\begin{remark} 
The objective of ML detection is to minimize the  objective function \eqref{eq:f_bpsk}, which is equivalent to minimizing the following function 
\begin{eqnarray}
f(|s\rangle) = \sum_{l>k}^{N}  A_{k,l} s_k s_l  - \sum_{k=1}^{N}  b_k s_k, 
\end{eqnarray}
by omitting the constant values in \eqref{eq:f_bpsk}. Consequently, the problem Hamiltonian  $H_f$ of  the ML detection is equivalently transformed into
\begin{eqnarray}
\begin{aligned}
H_f =  \sum_{l>k}^N A_{k,l} \sigma_z^{(k)}  \sigma_z^{(l)} -  \sum_{k=1}^{N}   b_{k} \sigma_z^{(k)}.  \label{eq:Hp_f_simp}
\end{aligned}
\end{eqnarray}
\end{remark} 
Given $H_f$ of \eqref{eq:Hp_f_simp},  the QAOA prepares a parameterized unitary operator in terms of $H_f$ depending on an angle $\gamma$ as follows:
\begin{subequations}
\begin{align}
U(H_f,\gamma) & = e^{-i\gamma H_f} \label{eq:U_Hf1}\\
& = \prod_{k=1}^{N} \prod_{l>k}^{N-1} e^{-i \gamma A_{k,l} \sigma_z^{(k)}  \sigma_z^{(l)} }  e^{ i \gamma b_k \sigma_z^{(k)} } \label{eq:U_Hf2}\\
&=  \prod_{k=1}^{N} \prod_{l>k}^{N-1} U(A_{k,l},\gamma)  U(b_{k},\gamma), \label{eq:U_Hf3}
\end{align}
\end{subequations}
with $\gamma \in [0,2\pi]$. Note that  since  all terms in $H_f$ of \eqref{eq:Hp_f} are diagonal in the computational basis,  they commute with each other.  As a result, $U(H_f,\gamma)$ can be written as a sequence of unitary operators formulated  in \eqref{eq:U_Hf2} and \eqref{eq:U_Hf3},  where the  unitary operators  are  $U(A_{k,l},\gamma) = e^{-i \gamma A_{k,l} \sigma_z^{(k)}  \sigma_z^{(l)} } $ and  $U(b_{k},\gamma) = e^{ i \gamma b_k \sigma_z^{(k)} }$, respectively.  This means that  implementing $U(H_f,\gamma) $ requires $(N+1)N/2$ unitary gates.  
Based on the initial Hamiltonian $H_B$ of \eqref{eq:H_B},  let us  define the parameterized unitary operator $U(H_B,\beta)$ depending on the angle $\beta$ as 
\begin{eqnarray}
\begin{aligned}
U(H_B,\beta) = e^{-i \beta H_B} = \prod_{k=0}^{N-1} e^{-i \beta \sigma_x^{(k)}},  \label{eq:U_HB2}
\end{aligned}
\end{eqnarray}
where we have $\beta \in [0, \pi]$.  

Following the VQE principle,  the QAOA creates a parameterized state in terms of $\gamma$ and $\beta$ on the quantum computer using single-qubit and entangling gates as follows. Given an integer $p$, there are $2p$ angles $\boldsymbol{\gamma} = [\gamma_1, \cdots, \gamma_p]$ and  $\boldsymbol{\beta} = [\beta_1, \cdots, \beta_p]$.   Specifically, in the QAOA, an angle-dependent quantum state is created by alternately applying Hamiltonians $H_f$ and $H_B$ in $p$ consecutive rounds, which can be expressed as  
\begin{eqnarray}
\begin{aligned}
|\psi_p( \boldsymbol{\gamma}, \boldsymbol{\beta} )\rangle &= U(H_B,\beta_p) U(H_f,\gamma_p)  \cdots U(H_B,\beta_1)  U(H_f,\gamma_1)  | \psi(0)\rangle.
\end{aligned}
\end{eqnarray}
Another important component of the QAOA is the computation of  the expectation value of $H_f$ in the state of $|\psi_p( \boldsymbol{\gamma}, \boldsymbol{\beta} )\rangle$, which can be expressed as 
 \begin{eqnarray}
 \begin{aligned} \label{eq:F_p}
 F_p &= \langle \psi_p^{\dagger}( \boldsymbol{\gamma}, \boldsymbol{\beta} )| H_f | \psi_p( \boldsymbol{\gamma}, \boldsymbol{\beta} ) \rangle \\
 & =\langle \psi(0) |U^{\dagger}(H_f,\gamma_p)    \cdots  U(H_f,\gamma_1) H_f U(H_B,\beta_p)  \cdots U(H_f,\gamma_1)  |\psi(0)\rangle,
 \end{aligned}
 \end{eqnarray}
where $\dagger$ denotes the conjugate transpose.
Since the expectation value of $F_p$ relies on the parameterized state $\psi_p^{\dagger}( \boldsymbol{\gamma}, \boldsymbol{\beta} )$, the goal of the QAOA is to approximate the optimal $\boldsymbol{\gamma}^*$ and $\boldsymbol{\beta}^*$ that satisfies 
\begin{eqnarray}
\begin{aligned}
\boldsymbol{\gamma}^*, \boldsymbol{\beta}^* = \mathrm{arg} \min F_p(\boldsymbol{\gamma}, \boldsymbol{\beta} ). \label{eq:opt_Fp}
\end{aligned}
\end{eqnarray}
Specifically, in QAOA, the gate parameters $\boldsymbol{\gamma}$ and $\boldsymbol{\beta}$ are designed on the classical computer by optimizing the expectation values of $H_f$, which is obtained by measuring the state of the quantum system.
% Repeating the above procedure altogether, the optimal solution can be approximately achieved.  
%Therefore, QAOA generally can be implemented via shallow-depth circuits as a consequence of  this hybrid approach.  
Fig. \ref{fig:qaoa_diag} illustrates two schemes of realizing QAOA.  Fig. \ref{fig:qaoa_diag_vqe} illustrates the main steps of realizing QAOA using hybrid quantum-classical approaches based on VQE principles, which is suitable for the case,  where the depth of the quantum circuits is excessive  for  NISQ devices or the number of qubits is too high for computations to be carried out  by a classical computer.  Explicitly, the parameterized trial states $\psi(\boldsymbol{\gamma},\boldsymbol{\beta})$ are created on the quantum device, starting from the initial state $\psi(0)$.  The output from the quantum devices is the measurement of the expectation values of the state $\psi(\boldsymbol{\gamma},\boldsymbol{\beta})$, which are fed into the classical computer for updating  $\boldsymbol{\gamma}$ and $\boldsymbol{\beta}$ by the classical optimizer. The new  parameters $\boldsymbol{\gamma}$ and $\boldsymbol{\beta}$ are then fed back to the quantum device to adjust the system state.  The algorithm terminates,  when the minimized value of $F_p$ is reached.  By contrast, when  the circuit is shallow and when the number of qubits in terms of the problem Hamiltonian $H_f$  is not too high, the expectation value $F_p$ can be calculated classically, as illustrated in Fig. \ref{fig:qaoa_diag_clas}. To this end,  the analytical expression of  $F_1$ can be derived, which is given in {\bf Proposition \ref{pro:qaoa_part}}.
%In particular, 
% and   measuring the expectation values.
%The QAOA in principle consists of four steps as shown in 
   
\begin{proposition} \label{pro:qaoa_part}
For the QAOA associated with $p=1$,   the expectation value $F_1$,  depending on a pair of parameters $(\gamma, \beta)$ omitting the subscript for notational simplicity, can be calculated analytically, which is given as follows:
\begin{enumerate}
\item When $N=1$, we arrive at:
\begin{eqnarray}
\begin{aligned}
F_1 = b \sin(2\beta) \sin(2b \gamma). \label{eq:F1_N1}
\end{aligned}
\end{eqnarray}
\item When $N=2$, we have 
\begin{eqnarray}
\begin{aligned}
F_1 = A_{1,2} f_{1,2}+\sum_{k=1}^{2} g_k, \label{eq:F1_N2}
\end{aligned} 
\end{eqnarray}
where 
\begin{subequations}
\begin{align*}
f_{1,2} &= \langle ++|U^{\dagger}(A_{1,2},\gamma)U^{\dagger}(b_1,\gamma)U^{\dagger}(b_2,\gamma) U^{\dagger}(\beta^{(1)})\\& \qquad U^{\dagger}(\beta^{(2)}) \sigma_z^{1}\sigma_z^{2} U(\beta^{(2)})U(\beta^{(1)})  U(b_2,\gamma) U(b_1,\gamma) U(A_{1,2},\gamma)|++\rangle,\\
g_k &= \langle ++|U^{\dagger}(A_{1,2},\gamma)U^{\dagger}(b_1,\gamma)U^{\dagger}(b_2,\gamma) U^{\dagger}(\beta^{(1)})   U^{\dagger}(\beta^{(2)}) \sigma_z^{k}U(\beta^{(2)})U(\beta^{(1)}) U(b_2,\gamma)U(b_1,\gamma)  U(A_{1,2},\gamma)|++\rangle,
\end{align*}
\end{subequations}
with $U(\beta^{(k)}) = e^{-i \beta\sigma_x^{(k)}}$ representing the unitary operator  acting on the $k$-th qubit. 
\item When $N\geq 3$, we have 
\begin{eqnarray}
\begin{aligned}
F_1 (\gamma,\beta) & =\sum_{l>k}A_{k,l} f_{k,l}   - \sum_{k}b_{k} g_k,  \label{eq:F1_N3}
\end{aligned}
\end{eqnarray}
where 
\begin{subequations}
\begin{align*}
f_{k,l} &= \langle +^N | U_{N}^{\dagger}(\gamma,\beta)  \cdots   U_{1}^{\dagger}(\gamma,\beta)  \sigma_z^{(k)}\sigma_z^{(l)} U_{1} (\gamma,\beta)  \cdots   U_{N} (\gamma,\beta) |+^N\rangle, ~~\text{and} \\
g_{k} &= \langle+^3| U^{\dagger}(A_{k',k},\gamma)  U^{\dagger}(A_{k,k''},\gamma)   U^{\dagger} (b_k,\gamma)  U^{\dagger}(1,\beta) \sigma_z^{k}  U(1,\beta) U(b_k,\gamma) U(A_{k,k''},\gamma)   \\& \qquad U^{\dagger}(A_{k',k},\gamma)  \sigma_z^{(k)} |+^3\rangle,\quad k = 1,2, 
\end{align*}
\end{subequations}
 with  $U_{k}(\gamma,\beta) = \prod_{l>k} U(A_{k,l},\gamma)  U(b_{k},\gamma) U(k,\beta) $.
\end{enumerate}

\begin{proof}
See Appendix~D.
\end{proof}
\end{proposition}

\section{Complexity Analysis}
\label{sec:comanal}
In this section, we provide  the computational complexity analysis of the QAOA for solvong the ML detection problem of interest,  when the analytical expression of $F_p$ is given.  In this context, the maximization of $F_p$ can be solved by classical  solvers such as COBYLA \cite{powell2007view} via  linear approximation techniques. Then, the quantum circuits of QAOA are  constructed based on the resultant $\boldsymbol{\gamma}^*$ and $\boldsymbol{\beta}^*$.
 The complexity of  QAOA in terms of its quantum implementation directly depends on the number of gates required.  Observe from \eqref{eq:U_Hf3} and \eqref{eq:U_HB2} that implementing $U(H_f, \gamma)$ and $U(H_B, \beta)$ requires $(N+3)N/2$ unitary gates in total.  Furthermore, the preparation of the initial state $\psi_0$ requires $N$ Hadamard gates, while the circuits of QAOA associated with the depth of $p$ involves $(N+5)Np/2$ quantum gates altogether.  In the following, we discuss the computational complexity of the classical solver of \eqref{eq:opt_Fp} as well as  the memory requirement of simulating the QAOA on a classical computer. 

\subsubsection{Computational complexity of the classical optimizer}
%\label{sec:sub-csol}
The optimization of the expectation value $F_p$ of \eqref{eq:opt_Fp} involves $2p$ variables in total.  Since we use  COBYLA as the classical solver for optimization of \eqref{eq:opt_Fp},  it requires $O(m^2)$ function evaluations for each iteration \cite{powell2007view}.   Note that $m$ is the number of interpolation points, which is given by  $m = \frac{1}{2}(2p+1)(2p+2) = (p+1)(2p+1) = 2p^2+3p+1$. The number of function evaluations for each iteration is thereby $O(p^4)$.  From \eqref{eq:F_p}  and  {\bf Proposition \ref{pro:qaoa_part}},  we see that computing $F_p$ relies on matrix dot  products and matrix exponential operations. In the worst case,  all of the unitary operators are associated with a $2^N \times 2^N$  matrix, hence  the computational  complexities of  matrix exponential evaluations and matrix dot products are on the order of  $O(2^{3N})$  and $O(2^{2N})$, respectively.  As seen from \eqref{eq:F_p},  the computation  of $F_p$ requires $4p$ matrix exponential operations and $4p+1$ matrix dot product operations. As a result, the  computational complexity of each evaluation $F_p$  is given by
\begin{equation}
\begin{split}
O(F_p) &= 4pO(8^{N})+(4p+1)O(4^{N}) \\
&= O(p8^{N}).
\end{split}
\end{equation}
%The complexity of O(\text{COBYLA-per-iteration}) = O(p^4*p8^N) = O(p^58^N)$
Therefore, the total computational complexity of solving problem \eqref{eq:opt_Fp}  is  on the order of $ O(Ip^58^N)$, where $I$ is the number of the iterations.  This indicates that the computational complexity of the classical solver grows exponentially with $N$ even if $p=1$. 

\subsubsection{Memory requirement of simulating the QAOA classically} 
%\label{sec:sub-sc}
When evaluating  $F_p$  classically, the classical computer will have to store both  the quantum states and  the unitary operators in terms of $U(H_f,\gamma)$ and $U(H_B, \beta)$.  Specifically, a state of $N$ qubits is characterized by a  $2^N$-element complex vector, which requires $2^N$ complex numbers and $2^{N+1}$ floating-point numbers. In a classical computer, a floating-point number is typically represented by four bytes in single-precision floating-point format.  Correspondingly, we have to use $2^{3+N}$ bytes for  storing a quantum state of $N$ qubits.   Furthermore,  a unitary matrix contains  $2^{2N}$ complex numbers, which thereby occupies $2^{2N+3}$ bytes in the memory.  Therefore, the  amount of memory required  is dominated by storing the unitary operators. For instance, if we have a computer equipped with 16G RAM, the  number of qubits that can be handled is at most $N=15$. On the other hand, based  on the VQE principle, evaluating the expectation value containing the parameterized state and the unitary operators is performed by the quantum devices, while the classical computer  only needs to store the current quantum state that is simulating at the quantum device.  In this context, the amount of memory required depends on the storage required by  a single quantum state. Correspondingly, the maximum number of qubits that can be simulated using 16G RAM is $N = 30$.

\section{Simulation results}
\label{sec:sim}

In this section, we first characterize the eigenvalues of the problem Hamiltonian for the ML detection. Then, we visualize the expectation values $F_1$ in terms of  different number of qubits. Finally, we quantify the performance of QAOA in the ML detection of MIMO systems using computer simulations.  As for performing quantum computing, Qiskit Aer \cite{Qiskit} is used for implementing the noise-free simulations of QAOA circuits.
%where  the expectation values $F_1$ can be computed using Mathematica and the BER performance of QAOA based ML detection is simulated using python 
%In our first simulations, we plot the landscape of expectation values based on the analytical expressions in {\bf Proposition \ref{pro:qaoa_part}}. Afterwards, the BER performance 

\begin{figure} [htp]
\centering
\subfigure[Different eigenvalues]{
\includegraphics[clip, height = 2.0in,width = 0.47\textwidth]{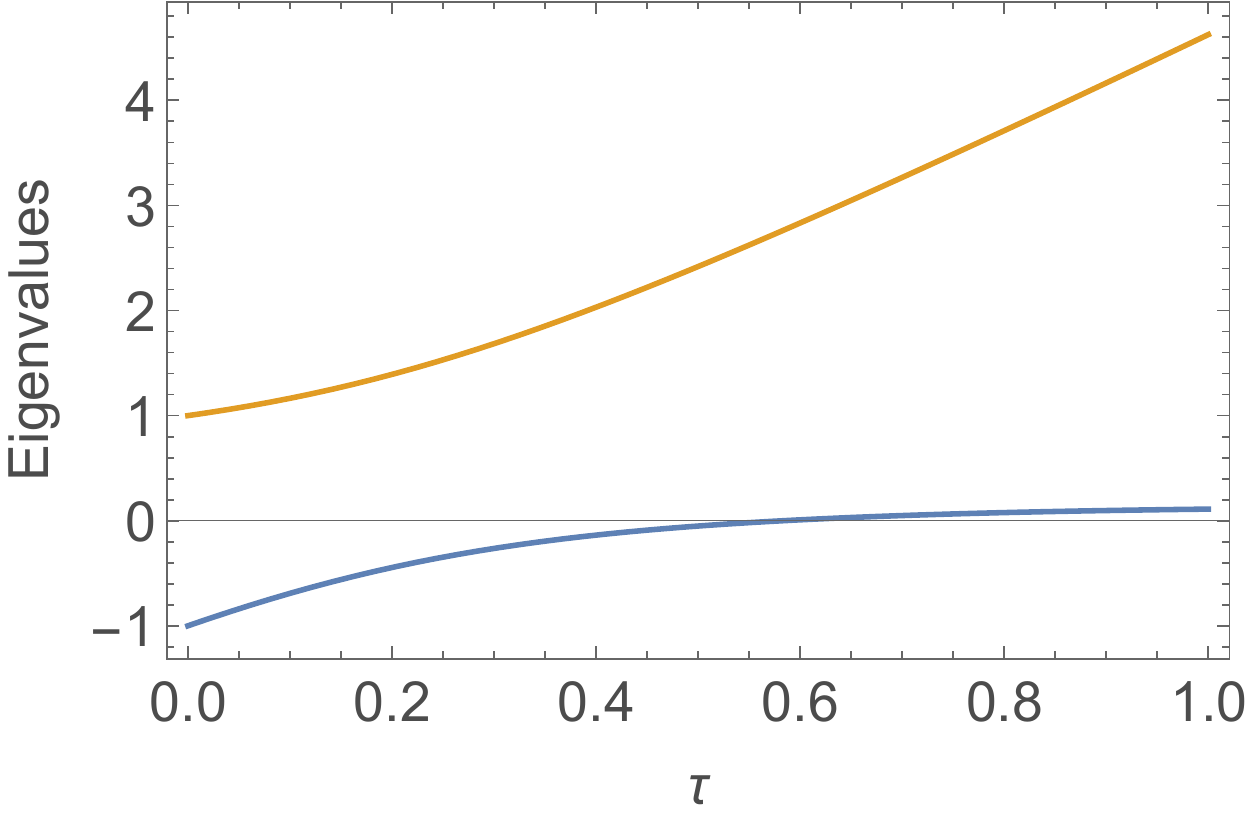}
 \label{fig:E_one_qbit}
}
\subfigure[Gap ]{
\includegraphics[clip, width = 0.45\textwidth]{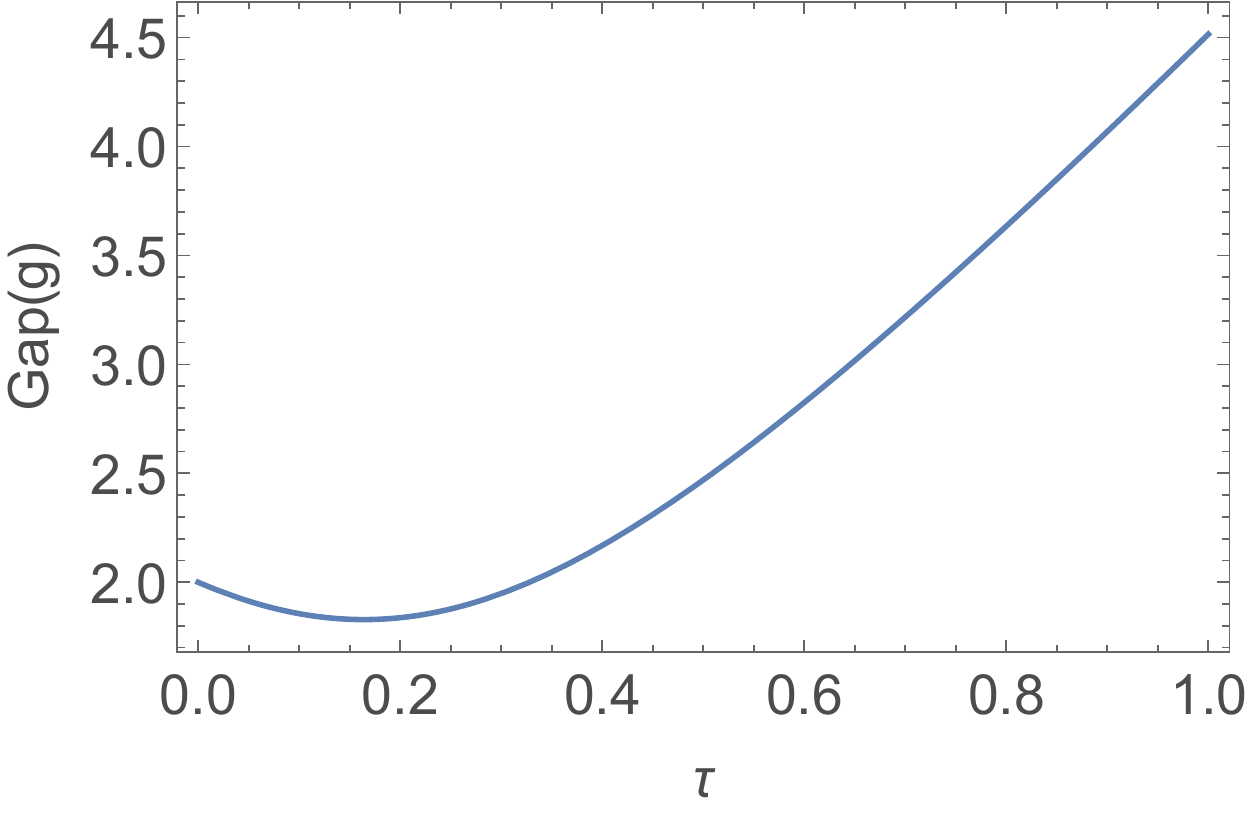}
 \label{fig:gap_oneqbit}
}
  \caption{The two eigenvalues of $\tilde{H}(\tau)$ for a single-qubit example. }
 \label{fig:ex_onequbit}
   \vspace{-1.0em}
\end{figure}

Fig. \ref{fig:ex_onequbit} depicts the evolution of the eigenvalues for a single-qubit ML detection problem, where  we consider a concrete example  associated with a random channel coefficient of $h =  1.2416$, $z = 0.3323$ and the transmit signal $s = 1$.  The two eigenvalues of $\tilde{H}(s)$ and the corresponding gap ($g$) of \eqref{eq:min_gap} are  plotted in Fig. \ref{fig:E_one_qbit} and in Fig. \ref{fig:gap_oneqbit}, respectively. We  find that the minimum gap $g = 1.94 $ is not small and thus it is possible for driving the system to  evolve smoothly from $\psi(0) = |x = 0 \rangle $ to $| z = 0\rangle$.   

\begin{figure} [htp]
\centering
\subfigure[Two qubits]{
\includegraphics[clip, height = 2.0in,width = 0.47\textwidth]{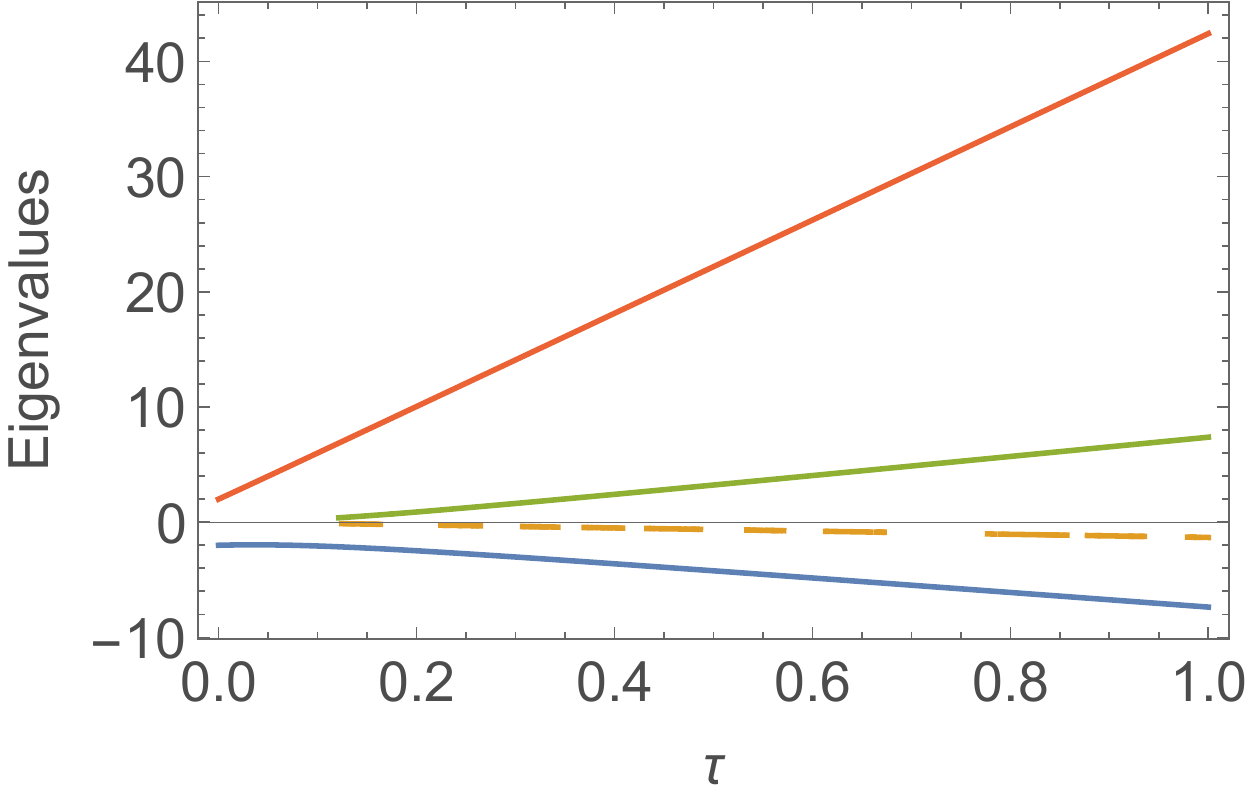}
 \label{fig:E_2_qbit}
}
\subfigure[Three qubits ]{
\includegraphics[clip, width = 0.4\textwidth]{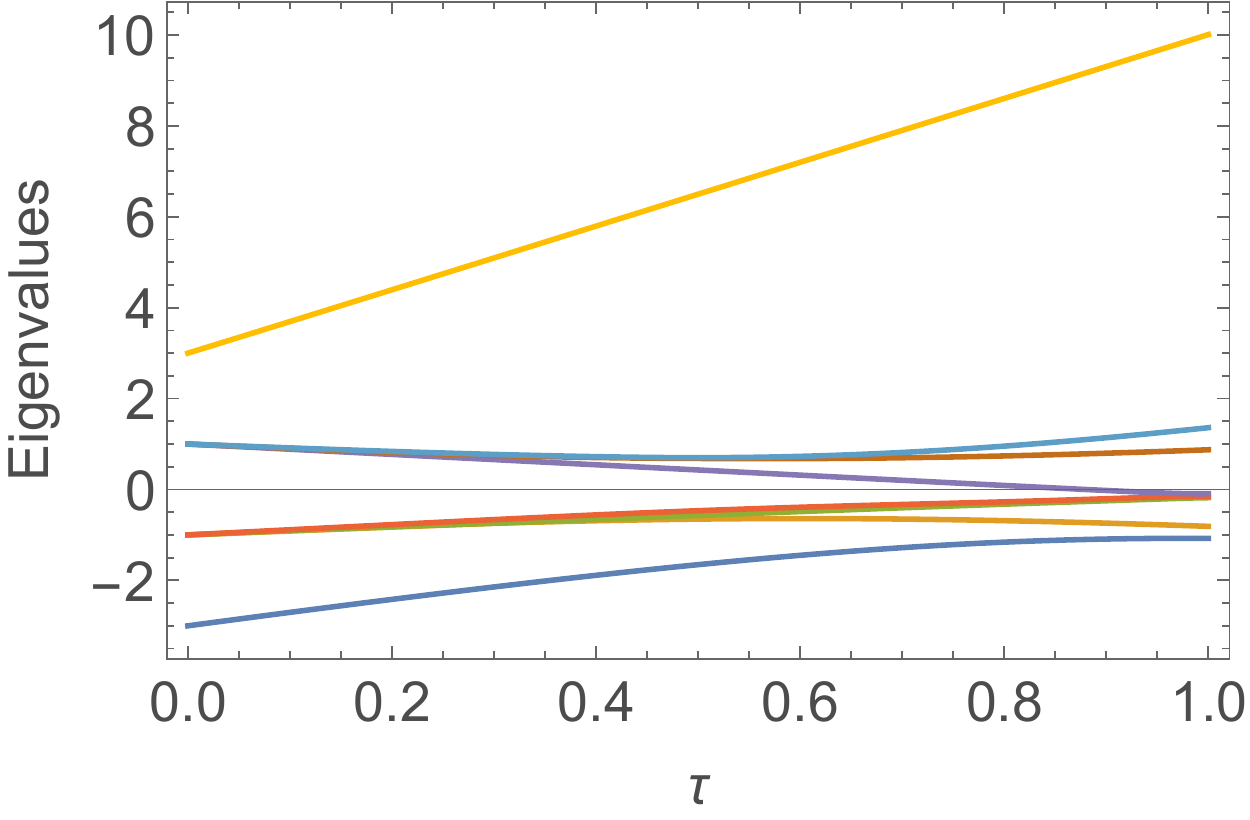}
 \label{fig:E_3_qbit}
}
  \caption{Different eigenvalues of $\tilde{H}(\tau)$ for two and three-qubit examples. }
 \label{fig:E_2_3_qbit}
   \vspace{-1.0em}
\end{figure}

Furthermore, the evolutions of eigenvalues for the two-qubit and three-qubit  ML detection problems are portrayed in Fig, \ref{fig:E_2_3_qbit}.  In Fig. \ref{fig:E_2_qbit}, we plot the four eigenvalues of $\tilde{H}(\tau)$ for a group of random system parameters, where the values are set as  $\mathbf{H} = [[1.2416, -0.1741],[0.3323, -0.0804]]$, $n = [-1.5130, 0.3212]$ and $\mathbf{s} = [-1,+1]$. Then, the minimum gap of \eqref{eq:min_gap} can be found as $g = 1.93$, which indicates that the optimal solution may indeed be obtained by smoothly evolving the system from the initial ground state to the final ground state, when $T \gg 1/g^2$.  Fig. \ref{fig:E_3_qbit} illustrates the eigenvalues of $\tilde{H}(\tau)$ for a three-qubit ML problem, where we have 
the channel matrix of $\mathbf{H} = [[1.24155, -0.174105, 0.332349],$ $ [-0.080418, -1.51301, 
  0.321184], [-1.7771, 1.55398, 0.23342]]$, the noise of $\mathbf{n} = [-1.703, -1.77439, 1.34985]$ and the transmit signal sequence is $\mathbf{x} = [-1, 1, 1]$.
We can see  that the minimum gap is non-zero, which indicates   convergence to the optimal solution. Furthermore, we see that there are  some overlapping eigenvalues in Fig. \ref{fig:E_2_3_qbit}, which arises from the symmetry of $\tilde{H}(\tau)$ \cite{farhi2000quantum}.

Fig. \ref{fig:ex_F1}  shows the landscape of expectation values for a depth of $p=1$ circuits with two parameters $\gamma_1$ and $\beta_1$, where the expectation values of $F_1(\gamma_1,\beta_1)$ for $N=1,2$ and $3$ are plotted in  \eqref{eq:F1_N1}, \eqref{eq:F1_N2} and \eqref{eq:F1_N3}, respectively.  Since  the analytical expressions of $F_1(\gamma_1,\beta_1)$ for $N=1,2$ and $3$ are given in {\bf Proposition \ref{pro:qaoa_part}}, we plot their figures using Mathematica.  
Furthermore, the values of the channel coefficients and the noise  are the same as   in the corresponding numerical examples of   Fig. \ref{fig:ex_onequbit} and Fig. \ref{fig:E_2_3_qbit}.  As $F_p$ is an even function, i.e. $F_p(\gamma,\beta)  = F_p(-\gamma,-\beta) $, the graph of $F_p$ is symmetric with respect to the axis of the values $F_p$, which allows us to restrict the range of $\gamma$ and $\beta$ to $[0,\pi]$.  
 We also see that in Fig.  \ref{fig:ex_F1}  $F_1$ is non-convex in terms of $\gamma_1$ and $\beta_1$, since it has multiple local minima. Furthermore, the number of locally optimal points grows upon increasing both $N$ and  the depth  $p$ of the circuits, especially when the expectation values have to  be evaluated by sampling from the quantum circuit measurements. This poses challenges in  classical simulations.  
\begin{figure} [!t]
\centering
\subfigure[$N=1$]{
\includegraphics[ width = 0.45\textwidth]{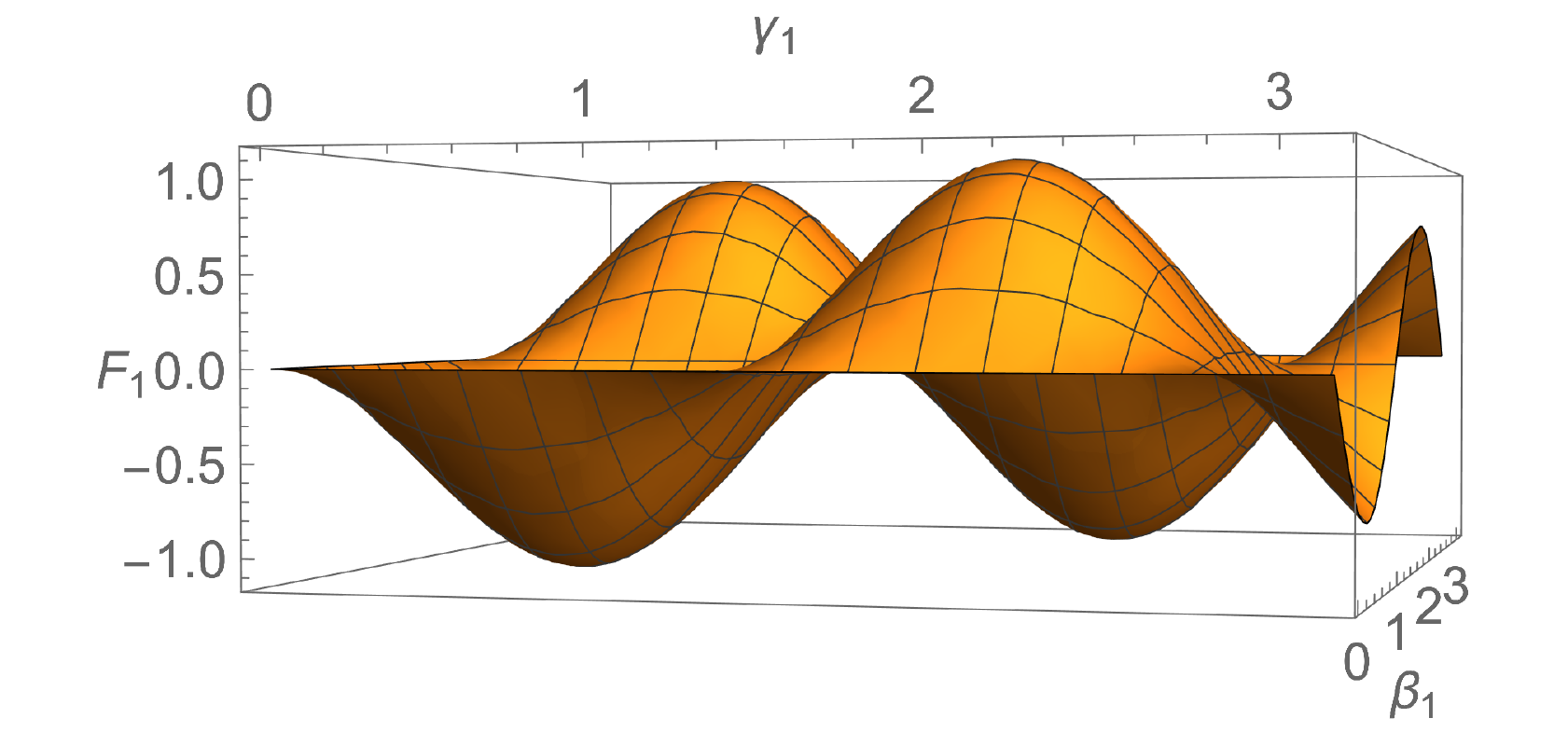}
\label{fig:ex_F1}
}
\subfigure[$N = 2$]{
\includegraphics[ width = 0.45\textwidth]{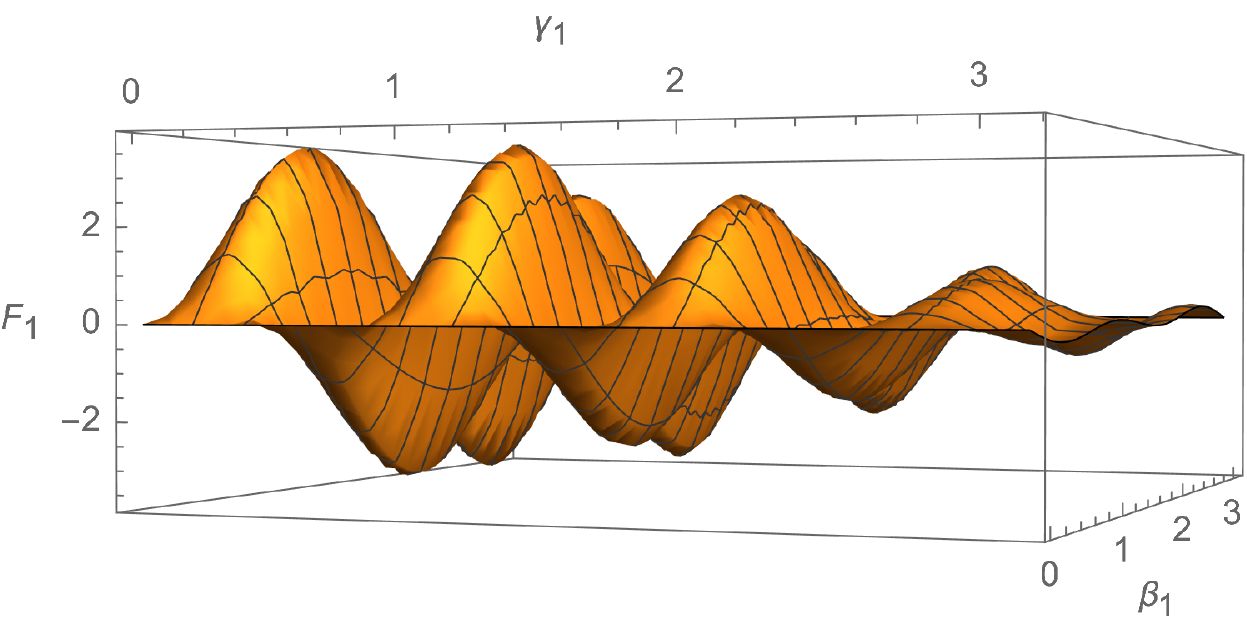}
\label{fig:ex_F2}
}
\subfigure[$N=3$]{
\includegraphics[ width = 0.45\textwidth]{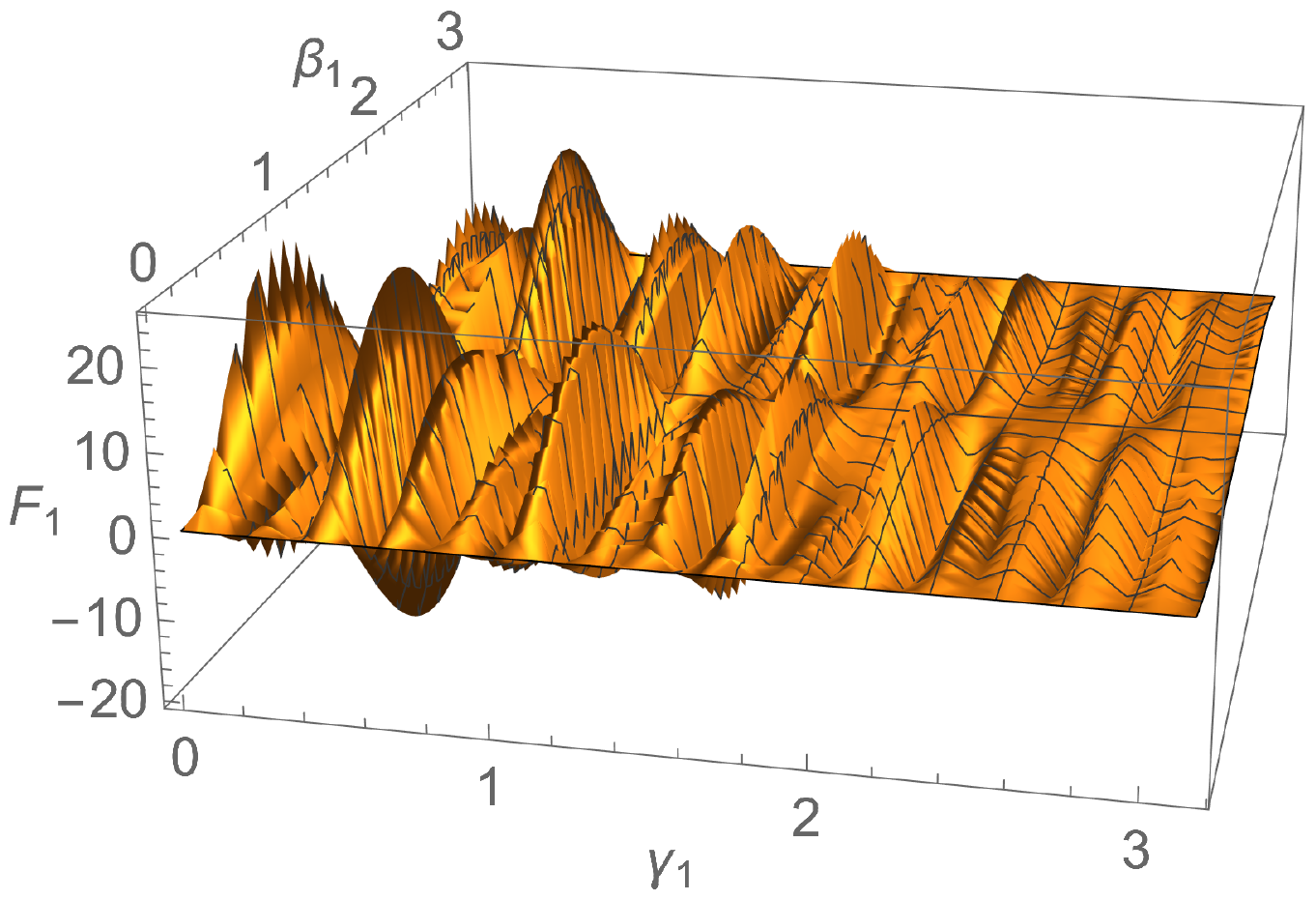}
\label{fig:ex_F3}
}
 \caption{The expectation value function $F_1(\gamma_1,\beta_1)$.}
 \label{fig:ex_F1}
   \vspace{-1.5em}
\end{figure}

\begin{figure} [!t]
\centering
\subfigure[$N=2$]{
\includegraphics[ width = 0.45\textwidth]{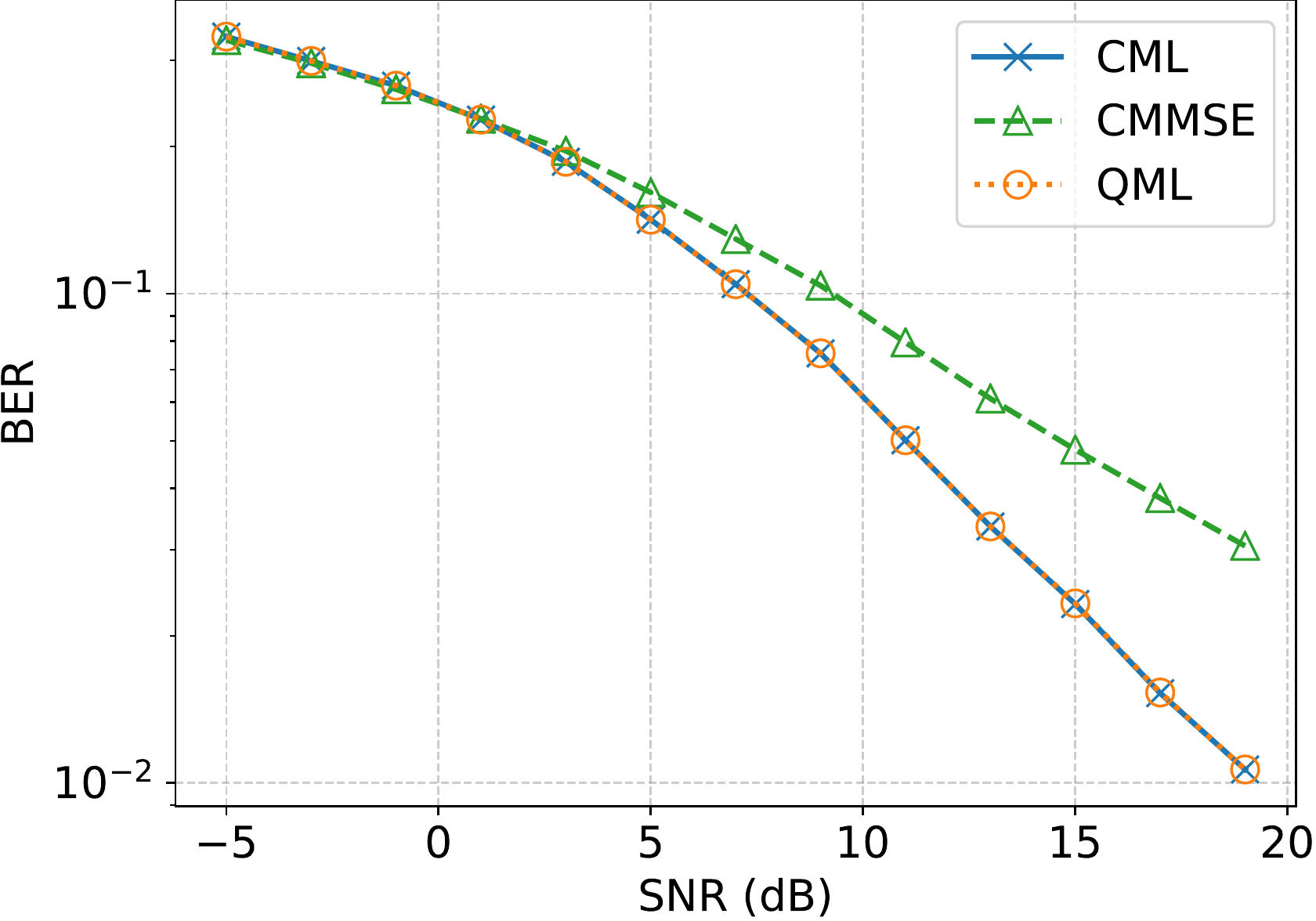}
\label{fig:ber_bpsk_N2}
}
\subfigure[$N = 3$]{
\includegraphics[ width = 0.45\textwidth]{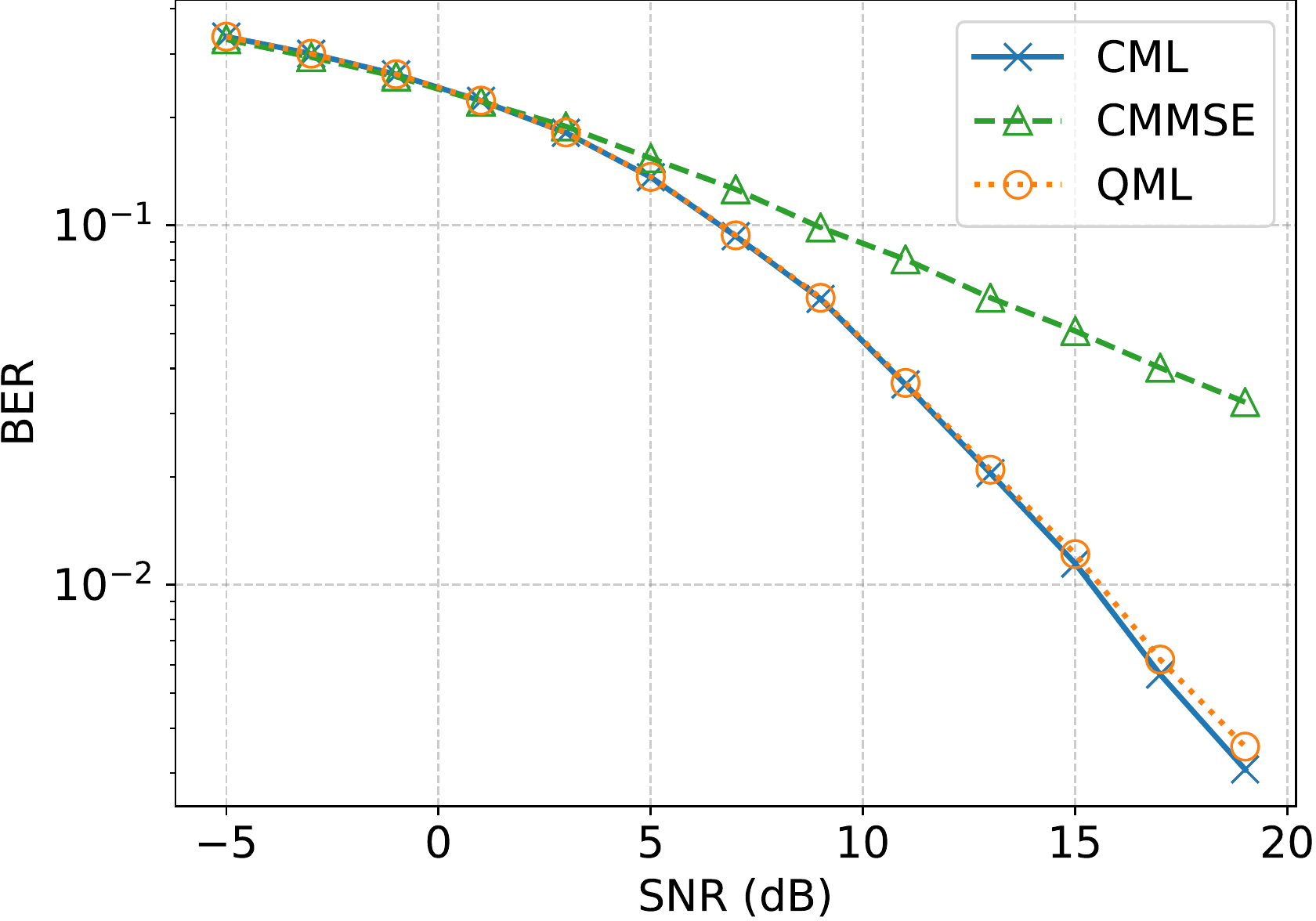}
\label{fig:ber_bpsk_N3}
}
\subfigure[$N=4$]{
\includegraphics[ width = 0.45\textwidth]{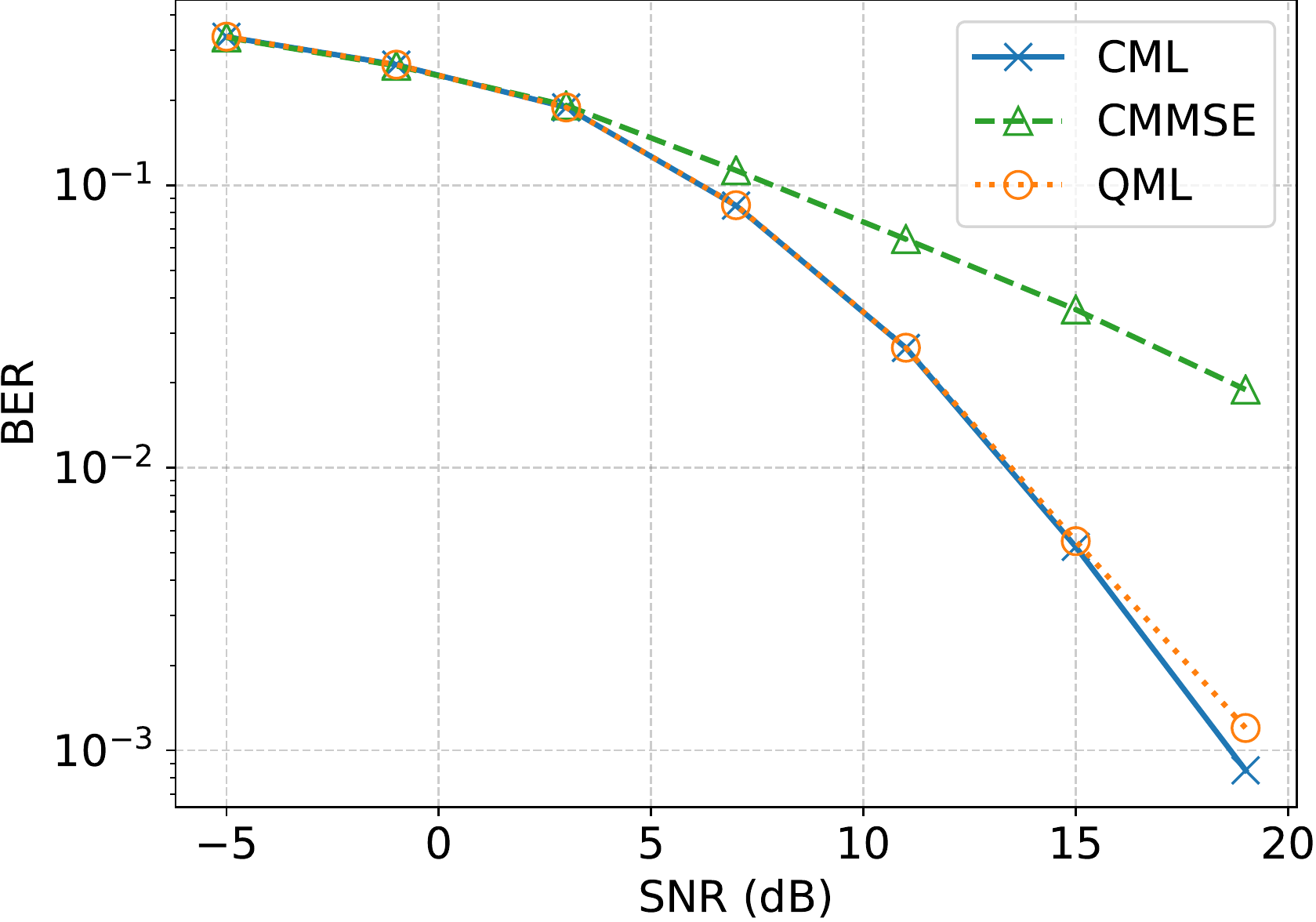}
\label{fig:ber_bpsk_N4}
}
 \caption{The BER performance of MIMO systems with different number of users.}
 \label{fig:ber_bpsk}
   \vspace{-1.5em}
\end{figure}

As discussed in Section \ref{sec:comanal}, simulating QAOA in classical computers requires excessive amount of memory and computational power. Therefore, we confine our simulations to the ML detection problem within 4 qubits.  In the simulations, the MIMO channel $\mathbf{H}$  and the noise $\mathbf{n}$ were chosen as independent and identically distributed, zero-mean, real-valued normal random variables, i.e. $\mathbf{H}\sim \mathcal{N}(0,1)$ and $\mathbf{n}\sim \mathcal{N}(0,1)$.  Furthermore,  for each SNR, we perform 40 000 Monte Carlo simulations for estimating the average probability of errors in detecting the message vector.  
 For benchmarking  the QAOA based ML (QML) detector, we consider a   pair of conventional detection methods: Classical ML (CML) and classical MMSE (CMMSE).  Fig. \ref{fig:ber_bpsk} illustrates the BER of  the three different detection methods  for  $N=2,~3$ and $4$, where we consider a MIMO system having the same  number of transmit and receive antennas, i.e. $M_t = M_r = N$.  Observe from Fig. \ref{fig:ber_bpsk} that the QAOA based ML detector approaches  the  BER of  classical ML detector,  and as expected both outperform the MMSE detector.  Furthermore, in  Fig. \ref{fig:ber_bpsk_N2}, we can see that the BER curve of QML  perfectly matches that of CML. However, in Fig. \ref{fig:ber_bpsk_N3} and Fig. \ref{fig:ber_bpsk_N4}, the BER of QML becomes slightly worse than that of CML in the high-SNR region. This is because the QML solution is  estimated statistically  relying on  the quantum circuit measurements.

\section{Conclusions}
\label{sec:conclu}
In this paper, we studied the performance of QAOA based ML detection problems, where we considered the ML detection of  binary symbols over a MIMO channel.    We first encoded the optimal solution of the ML detection problem into the ground state of a problem Hamiltonian and presented the energy evolution of the quantum system of interest. For level-1 QAOA, we derived the analytical expressions of QAOA and provided the energy landscape of QAOA that illustrates its symmetry vs.  the parameter values to be optimized. Finally, our simulation results revealed that QAOA based ML detection  is capable of approaching the  BER  of the classical ML detector, while both  outperform the  classical MMSE detector.

\numberwithin{equation}{section}
%\section*{Appendix~A: Proof of Proposition \ref{pro:C_measure}} \label{Appdx:C_measure}
\section*{Appendix~A: Proof of The Problem Hamiltonian $H_f$} \label{Appdx:A}
\renewcommand{\theequation}{A.\arabic{equation}}
\setcounter{equation}{0}

For mapping the variable  $z_k = \{0,1\}$ into $s_k \in \{-1,+1\}$, we define a bijection function $g(z_k)$:  $s_k = g(z_k) = 1 - 2z_k$\footnote{ Note that the mapping function $g(z_k)$ is usually not unique, any bijective function such that $g:\{0,1\}\rightarrow \{-1,1\}$ can be chosen as the mapping function such as $g(z_k) = (-1)^{z_k} $.}.  Correspondingly, the objective function of \eqref{eq:f_bpsk} can be equivalently reformulated as 
%\begin{eqnarray}
%\begin{aligned}
%f(|z\rangle) &= \sum_{l>k}^{N} 2 A_{k,l} (1 - 2z_k) (1 - 2z_l)  - \sum_{k=1}^{N} 2 b_k (1 - 2z_k) + c + \sum_{k=1}^N A_{k,l}. \label{eq:f_bpsk_z}
%\end{aligned}
%\end{eqnarray}
\begin{dmath}
f(|z\rangle) = \sum_{l>k}^{N} 2 A_{k,l} (1 - 2z_k) (1 - 2z_l)  - \sum_{k=1}^{N} 2 b_k (1 - 2z_k) + c + \sum_{k=1}^N A_{k,l}. \label{eq:f_bpsk_z}
\end{dmath}

Furthermore, we can rewrite \eqref{eq:pauli_z} as $\sigma_z^{(k)} |z_k \rangle = (1 - 2 z_k) |z_k\rangle = s_k|z_k\rangle$, which indicates that the binary variables $s_k$ is mapped onto the eigenvalues of the Pauli-Z operator. Now we show that $H_f|z\rangle = f(|z\rangle) |z\rangle$ in \eqref{Outre3.part2}.
{\small
\begin{eqnarray} \label{Outre3.part2}
\begin{aligned}
H_f|z\rangle ={ }&  \left[ \sum_{l>k}^N 2A_{k,l} \sigma_z^{(k)}  \sigma_z^{(l)} -  \sum_{k=1}^{N}  2 b_{k} \sigma_z^{(k)}  + c + \sum_{k=1}^N A_{k,k}\right] 
 |z_1,\cdots, z_n\rangle \\
={ }& \sum_{l>k}^N 2A_{k,l} \sigma_z^{(k)}  \sigma_z^{(l)} |z_1,\cdots, z_n\rangle -  \sum_{k=1}^{N}  2 b_{k} \sigma_z^{(k)} |z_1,\cdots, z_n\rangle 
+ c + \sum_{k=1}^N A_{k,k} \\
 ={ }& \sum_{l>k}^N 2A_{k,l} (1 - 2 z_k)(1 - 2z_l) |z_1,\cdots, z_n\rangle -  \sum_{k=1}^{N}  2 b_{k} (1 - 2 z_k) |z_1,\cdots, z_n\rangle + c + \sum_{k=1}^N A_{k,k}  \\
 ={ }& \left[\sum_{l>k}^N 2A_{k,l} (1 - 2 z_k)(1 - 2z_l)  -  \sum_{k=1}^{N}  2 b_{k} (1 - 2 z_k)  + c + \sum_{k=1}^N A_{k,k} \right] |z_1,\cdots, z_n\rangle \\
& = f(|z\rangle) |z\rangle.
\end{aligned}
\end{eqnarray}}
\iffalse
\newcounter{mytempeqncnt}
%\hrulefill %这里有一条线，如果你想要
\begin{figure*}[ht]%公式位置按图放置调整
\normalsize
\setcounter{mytempeqncnt}{\value{equation}}
\setcounter{equation}{1} %用于设定当前长公式编号，如此处为1，则编号为2
{\small
\begin{eqnarray} \label{Outre3.part2}
\begin{aligned}
H_f|z\rangle ={ }&  \left[ \sum_{l>k}^N 2A_{k,l} \sigma_z^{(k)}  \sigma_z^{(l)} -  \sum_{k=1}^{N}  2 b_{k} \sigma_z^{(k)}  + c + \sum_{k=1}^N A_{k,k}\right] 
 |z_1,\cdots, z_n\rangle \\
={ }& \sum_{l>k}^N 2A_{k,l} \sigma_z^{(k)}  \sigma_z^{(l)} |z_1,\cdots, z_n\rangle -  \sum_{k=1}^{N}  2 b_{k} \sigma_z^{(k)} |z_1,\cdots, z_n\rangle 
+ c + \sum_{k=1}^N A_{k,k} \\
 ={ }& \sum_{l>k}^N 2A_{k,l} (1 - 2 z_k)(1 - 2z_l) |z_1,\cdots, z_n\rangle -  \sum_{k=1}^{N}  2 b_{k} (1 - 2 z_k) |z_1,\cdots, z_n\rangle + c + \sum_{k=1}^N A_{k,k}  \\
 ={ }& \left[\sum_{l>k}^N 2A_{k,l} (1 - 2 z_k)(1 - 2z_l)  -  \sum_{k=1}^{N}  2 b_{k} (1 - 2 z_k)  + c + \sum_{k=1}^N A_{k,k} \right] |z_1,\cdots, z_n\rangle \\
& = f(|z\rangle) |z\rangle.
\end{aligned}
\end{eqnarray}}
%\setcounter{equation}{\value{mytempeqncnt}}
\hrulefill % %这里有一条线，如果你想要
\vspace*{-10pt} %留空白，可自己调整
\end{figure*}
%长公式结束
\setcounter{equation}{2}%结束后别忘了用此命令调整下一个公式编号
\fi
Therefore, we can see that the objective values of the ML detector is mapped onto the eigenvalues of the problem Hamiltonian $H_f$, where each bit string $|z\rangle$ associated with a variable vector $|s_1,\cdots,s_N\rangle$.

\numberwithin{equation}{section}
\section*{Appendix~B: Proof of {\bf Proposition \ref{pro:mu_mimo}}  } \label{Appdx:B}
\renewcommand{\theequation}{B.\arabic{equation}}
\setcounter{equation}{0}

Consider a multi-user SISO system, where $K$ users cooperatively  transmit independent symbols to a receiver. The receiver observes the sum of the modulated signals contaminated by the noise as follows
\begin{eqnarray}
y = \sum_{k=1}^K h_k s_k + n = \mathbf{h}^T \mathbf{s} + n, 
\end{eqnarray}
where $\mathbf{h} = [h_1, \cdots,h_K]^T$ and $\mathbf{s} = [s_1,\cdots,s_K]^T$.   Therefore, the objective function of our  ML detection problem becomes 
\begin{eqnarray}
\begin{aligned} \label{eq:siso}
f(\mathbf{s}) &= \mathbf{s}^T \mathbf{h}\mathbf{h}^T \mathbf{s} - 2 y \mathbf{h}^T\mathbf{s}+y^2 \\
&= \mathbf{s}^T \mathbf{A} \mathbf{s} - 2 \mathbf{b}\mathbf{s}+c \\
& = \sum_{l>k}^{K} 2A_{k,l} s_k s_l - \sum_{k=1}^{K} 2 b_k s_k + c + \sum_{k=1}^{K} A_{k,k},
\end{aligned}
\end{eqnarray} 
where $\mathbf{A} = \mathbf{h}\mathbf{h}^T$, $\mathbf{b} = y \mathbf{h}^T$, $c = y^2$  and $\mathbf{s} \in \{-1,+1\}^{K}$.    Furthermore,  $A^T = A$ is used to in the last step of \eqref{eq:siso}.  Following the form of \eqref{eq:Hp_f}, we have 
\begin{eqnarray}
H_f = \sum_{l>k}  A_{k,l} \sigma_z^{(k)}\sigma_z^{(l)} -  b \sum_{k=1}^K \sigma_z^{(k)}. \label{eq:Hf_siso}
\end{eqnarray}
Note that the constant terms  and the coefficient $2$ are omitted in \eqref{eq:Hf_siso}, since they do not affect the optimal solution of the original problem. From \eqref{eq:Hf_siso}, we can see that the number of qubits required is equal to the number of users for multi-user SISO systems.
Now we  extend the signal model to the corresponding multi-user MIMO system, which  can be expressed as: 
\begin{eqnarray}
\mathbf{y} = \sum_{k=1}^M \mathbf{H}_k \mathbf{s}_k + \mathbf{n} = \mathbf{H}^T \mathbf{s} + \mathbf{n},  \label{eq:mu_mino}
\end{eqnarray}
where $\mathbf{H} = [\mathbf{H}_1, \cdots,\mathbf{H}_k]^T$ and  $\mathbf{s} = [\mathbf{s}_1^T,\cdots,\mathbf{s}_K^T]^T$. Without loss of generality, we assume that $\mathbf{H}_k \in \mathcal{R}^{N \times N}$ and $\mathbf{s}_k \in \{-1,+1\}^N$.   As a result, we have $\mathbf{y}$ and $\mathbf{s}$ are $N \times 1$ and $NK \times 1$ vectors, respectively. 
Based on \eqref{eq:mu_mino},  the objective function of our  ML detection problem  can  be expressed as 
\begin{eqnarray}
\begin{aligned}
f(\mathbf{s}) &= \mathbf{s}^T \mathbf{H}\mathbf{H}^T \mathbf{s} - 2 \mathbf{y}^T \mathbf{H}^T\mathbf{s}+\mathbf{y}^T \mathbf{y} \\
& = \sum_{l>k}^{NK} 2A_{k,l} s_k s_l - \sum_{k=1}^{NK} 2 b_k s_k + c + \sum_{k=1}^{NK} A_{k,k},
\end{aligned}
\end{eqnarray}
where $\mathbf{A} = \mathbf{H}\mathbf{H}^T$ with $\mathbf{A}^T = \mathbf{A}$, $\mathbf{b} = \mathbf{y}^T \mathbf{H}^T$, $c = \mathbf{y}^T \mathbf{y}$  and $\mathbf{s} \in \{-1,+1\}^{NK}$.    Therefore, the Ising Hamiltonian of the ML detection to the corresponding multi-user MIMO system can be expressed as 
\begin{eqnarray}
\begin{aligned}
H_f =  \sum_{l>k}^{NK} A_{k,l} \sigma_z^{(k)} \sigma_z^{(l)}  - \sum_{k=1}^{NK} b_k \sigma_z^{(k)}, \label{eq:Hf_mu_mimo}
\end{aligned}
\end{eqnarray}
where the constant terms and the coefficient $2$ are also omitted. 
We can see from  \eqref{eq:Hf_mu_mimo} that the number of qubits required is $NK$ for  the ML detection  in a quantum computer.

\numberwithin{equation}{section}
\section*{Appendix~C: Matrix Form of Two-bit Example} \label{Appdx:C}
\renewcommand{\theequation}{C.\arabic{equation}}
\setcounter{equation}{0}

Here, the full matrix forms of  $H_f$, $H_B$ and $H_f$ for a two-qubit ML detection problem are given in \eqref{eq:Hp_f_twoqbit_full}, \eqref{eq:H_B_2qubit_full} and \eqref{eq:Hs_twoqbit_full}, respectively, where $c^{\prime} = c +  A_{1,1} + A_{2,2}$.

{\small
\begin{eqnarray}
\begin{aligned} \label{eq:Hp_f_twoqbit_full}
H_f &= 2  \sigma_z^{(1)}\sigma_z^{(2)} - 2(\sigma_z^{(1)}+\sigma_z^{(2)} ) +  c^{\prime} \\
&= {\footnotesize \begin{pmatrix}
c^{\prime} + 2A_{1,2} - 2b_1-2b_2& c^{\prime} & c^{\prime} & c^{\prime} \\
c^{\prime} & c^{\prime}-2A_{1,2} - 2b_1+2b_2 & c^{\prime}  & c^{\prime}  \\
c^{\prime}  & c^{\prime} &  c^{\prime} - 2A_{1,2} + 2b_1 - 2 b_2 & c^{\prime}  \\
c^{\prime}   & c^{\prime}   & c^{\prime}  & c^{\prime} + 2A_{1,2} + 2 b_1 +  2 b_2
\end{pmatrix}}.
\end{aligned}
\end{eqnarray}}
\iffalse
%\newcounter{mytempeqncnt}
%\hrulefill %这里有一条线，如果你想要
\begin{figure*}[ht]%公式位置按图放置调整
\normalsize
\setcounter{mytempeqncnt}{\value{equation}}
\setcounter{equation}{0} %用于设定当前长公式编号，如此处为1，则编号为2
{\small
\begin{eqnarray}
\begin{aligned} \label{eq:Hp_f_twoqbit_full}
H_f &= 2  \sigma_z^{(1)}\sigma_z^{(2)} - 2(\sigma_z^{(1)}+\sigma_z^{(2)} ) +  c^{\prime} \\
&= {\footnotesize \begin{pmatrix}
c^{\prime} + 2A_{1,2} - 2b_1-2b_2& c^{\prime} & c^{\prime} & c^{\prime} \\
c^{\prime} & c^{\prime}-2A_{1,2} - 2b_1+2b_2 & c^{\prime}  & c^{\prime}  \\
c^{\prime}  & c^{\prime} &  c^{\prime} - 2A_{1,2} + 2b_1 - 2 b_2 & c^{\prime}  \\
c^{\prime}   & c^{\prime}   & c^{\prime}  & c^{\prime} + 2A_{1,2} + 2 b_1 +  2 b_2
\end{pmatrix}}.
\end{aligned}
\end{eqnarray}}
%\setcounter{equation}{\value{mytempeqncnt}}
%\hrulefill % %这里有一条线，如果你想要
\vspace*{-10pt} %留空白，可自己调整
\end{figure*}
%长公式结束
\setcounter{equation}{1}%结束后别忘了用此命令调整下一个公式编号
\fi
%where $c^{\prime} = c +  A_{1,1} + A_{2,2}$.
%Furthermore, we take $H_B$  from the form of \eqref{eq:H_B}, with $N=2$, 
\begin{eqnarray}
\begin{aligned}
\label{eq:H_B_2qubit_full}
H_B &= \sigma_x^{(1)}+\sigma_x^{(2)} 
= {\footnotesize \begin{pmatrix}
0 & 1 & 1 & 0 \\
1 & 0 & 0 & 1 \\
1 & 0 & 0 & 1 \\
0 & 1 & 1 & 0 \\
\end{pmatrix}}.
\end{aligned}
\end{eqnarray}

{\small
\begin{eqnarray}
\begin{aligned} \label{eq:Hs_twoqbit_full}
\tilde{H}(\tau) &= (1-\tau) H_B + \tau H_f \\
&={\footnotesize \begin{pmatrix}
\tau(c^{\prime} + 2A_{1,2} - 2b_1-2b_2)& 1-\tau + \tau c^{\prime} & 1-\tau + \tau\tilde{c} & \tau c^{\prime} \\
1-\tau + \tau c^{\prime} & \tau( c^{\prime}-2A_{1,2} - 2b_1+2 b_2) & \tau c^{\prime}  &1-\tau + \tau c^{\prime}  \\
1-\tau + \tau c^{\prime}  & \tau c^{\prime} & \tau(  c^{\prime} - 2A_{1,2} + 2b_1 - 2 b_2) & 1-\tau + \tau c^{\prime}  \\
\tau c^{\prime}   & 1-\tau + \tau c^{\prime}   & 1- \tau + \tau  c^{\prime}  & \tau(c^{\prime} + 2A_{1,2} + 2 b_1 +  2 b_2)
\end{pmatrix}}.
\end{aligned}
\end{eqnarray}}

\iffalse
%\newcounter{mytempeqncnt}
%\hrulefill %这里有一条线，如果你想要
\begin{figure*}[!ht]%公式位置按图放置调整
\normalsize
\setcounter{mytempeqncnt}{\value{equation}}
\setcounter{equation}{2} %用于设定当前长公式编号，如此处为1，则编号为2
{\small
\begin{eqnarray}
\begin{aligned} \label{eq:Hs_twoqbit_full}
\tilde{H}(\tau) &= (1-\tau) H_B + \tau H_f \\
&={\footnotesize \begin{pmatrix}
\tau(c^{\prime} + 2A_{1,2} - 2b_1-2b_2)& 1-\tau + \tau c^{\prime} & 1-\tau + \tau\tilde{c} & \tau c^{\prime} \\
1-\tau + \tau c^{\prime} & \tau( c^{\prime}-2A_{1,2} - 2b_1+2 b_2) & \tau c^{\prime}  &1-\tau + \tau c^{\prime}  \\
1-\tau + \tau c^{\prime}  & \tau c^{\prime} & \tau(  c^{\prime} - 2A_{1,2} + 2b_1 - 2 b_2) & 1-\tau + \tau c^{\prime}  \\
\tau c^{\prime}   & 1-\tau + \tau c^{\prime}   & 1- \tau + \tau  c^{\prime}  & \tau(c^{\prime} + 2A_{1,2} + 2 b_1 +  2 b_2)
\end{pmatrix}}.
\end{aligned}
\end{eqnarray}}
\hrulefill % %这里有一条线，如果你想要
\vspace*{-10pt} %留空白，可自己调整
\end{figure*}
%长公式结束
\setcounter{equation}{3}%结束后别忘了用此命令调整下一个公式编号
\fi

\numberwithin{equation}{section}
%\section*{Appendix~A: Proof of Proposition \ref{pro:C_measure}} \label{Appdx:C_measure}
\section*{Appendix~D: Proof of Proposition \ref{pro:qaoa_part}} \label{Appdx:D}
\renewcommand{\theequation}{D.\arabic{equation}}
\setcounter{equation}{0}

For $p=1$ and $N=1$:    From \eqref{eq:Hp_f_simp} that omits the constants of \eqref{eq:Hp_oneqbit}, we arrive at the problem Hamiltonian in the form of  $H_f = -b\sigma_z$. Consequently, we have the unitary operators associated with $N=1$ as follows:
\begin{eqnarray}
\label{eq:U1}
\begin{aligned}
U(H_B,\beta_1) &= e^{-i \beta_1 \sigma_x }, ~\text{and} \\
U(H_f,\gamma_1) &= e^{i \gamma_1 b \sigma_z }.
\end{aligned}
\end{eqnarray}
The expectation value is therefore given by 
{\small
\begin{eqnarray}
\begin{aligned} \label{eq:F1_onequbit}
 &F_1(\gamma_1,\beta_1)   = -b \langle +| e^{-i \gamma_1 b \sigma_z }  e^{ i \beta_1 \sigma_x } \sigma_z   e^{ - i \beta_1 \sigma_x } e^{i \gamma_1 b \sigma_z } |+\rangle \\
 &=  -b \langle +| e^{-i \gamma_1 b \sigma_z } \Big( \cos(2\beta_1)\sigma_z + \sin(2\beta_1) \sigma_y  \Big)  e^{i \gamma_1 b \sigma_z } |+\rangle \\
  &= -b \langle +|  \cos(2\beta_1) + \sin(2\gamma_1 b) - \sin(2 \beta_1) \sin(2 \gamma_1 b) \sigma_x |+\rangle   \\
& = -b  \sin(2\beta_1)\sin(2 b \gamma_1), 
\end{aligned}
\end{eqnarray}}
where $\sigma_y$ represents the Pauli-Y gate. Furthermore, the following relationships are used in \eqref{eq:F1_onequbit}: $\sigma_z \sigma_x = -\sigma_x \sigma_z= i \sigma_y$, $ \sigma_x \sigma_y= - \sigma_x \sigma_y =  i\sigma_z$ and $\sigma_y \sigma_z = - \sigma_z \sigma_y  = i \sigma_x$.  Note that the double angle identities of $\cos 2x = \cos^2x - \sin^2 x$ and $\sin 2x = 2 \sin x \cos x$ are employed  in  \eqref{eq:F1_onequbit} as well. 

 Now we consider $p=1$ and $N=2$. The problem Hamiltonian then becomes: 
\begin{eqnarray}
H_f = A_{1,2} \sigma_z^{(1)}\sigma_z^{(2)} - \sum_{k=1}^2 b_k \sigma_z^{(k)}, \label{eq:H_f_2qubit}
\end{eqnarray}
From \eqref{eq:U_Hf3} and \eqref{eq:U_HB2}, we arrive at 
\begin{eqnarray} 
\begin{aligned}
%U(H_f) &=  e^{ i \gamma b_1 \sigma_z^{(1)}} e^{ i \gamma b_2 \sigma_z^{(2)}} , \\
%U(H_B) &= e^{-i \beta \sigma_x^{(1)}}  e^{-i \beta \sigma_x^{(2)}} 
U(H_f) &=  U(A_{1,2}, \gamma) U(b_1,\gamma) U(b_2,\gamma), \\
U(H_B) & = U(\beta^{(1)}) U(\beta^{(2)}) ,
\end{aligned}
\end{eqnarray}
where $U(A_{1,2}, \gamma) = e^{-i A_{1,2}  \gamma \sigma_z^{(1)}\sigma_z^{(2)}} $,  $U(b_k,\gamma) = e^{ i \gamma b_k \sigma_z^{(k)}}$. Furthermore,   $U(\beta^{(k)}) = e^{-i \beta \sigma_x^{(k)}}$ is the unitary operator acting on the $k$-th qubit.  By observing $H_f$ in \eqref{eq:H_f_2qubit},   we find that there are two individual components in terms of $A_{1,2}$ and $\{b_1 ,b_2\}$, respectively. As a result, the expectation values of the two components can be evaluated individually.  Let $f_{1,2} = \langle ++| U^{\dagger}(H_f,\gamma)U^{\dagger}(H_B,\beta) \sigma_z^{(1)}\sigma_z^{(2)}U(H_B,\beta) U(H_f,\gamma)$ and $g_k = \langle++|U^{\dagger}(H_f,\gamma)U^{\dagger}(H_B,\beta) \sigma_z^{(k)}U(H_B,\beta) U(H_f,\gamma)$. 
Correspondingly, we have the expectation value $F_1$ associated with $N=2$ as follows.
\begin{eqnarray}
F_1  = A_{1,2}f_{1,2} - b_1g_1 - b_2 g_2.
\end{eqnarray} 

For $N \ge 3$, the expectation value $F_1$ is the sum of the expectation values associated with the individual components of $H_f$ in \eqref{eq:Hp_f_simp}. Therefore,  we have 
 \begin{eqnarray}
 F_1 = \sum_{l>k} A_{k,l} f_{k,l} - \sum_{k}^{N} b_k g_k. 
 \end{eqnarray}
%  where $E = \{e = (k,l)| l>k, l,k = 1,\cdots,N\}$.
We first calculate $f_{k,l}$, which is given by 
\begin{eqnarray}
\begin{aligned}
f_{k,l} &= \langle \psi(0) |U^{\dagger}(H_f,\gamma)  U^{\dagger} (H_B,\beta) \sigma_z^{(k)}\sigma_z^{(l)}  U(H_B,\beta) U(H_f,\gamma)  |\psi(0)\rangle \\
& = \langle +^N | U_{N}^{\dagger} \cdots   U_{1}^{\dagger}  \sigma_z^{(k)}\sigma_z^{(l)} U_{1} \cdots   U_{N}  |+^N\rangle,
\end{aligned}
\end{eqnarray}
where $U_{k}(\gamma_1,\beta_1) = \prod_{l>k} U(A_{k,l},\gamma_1)  U(b_{k},\gamma) U(k,\beta_1) $.
For $g_k$,  all the unitary operators that do not intersect with the Pauli gate $\sigma_z^{k}$ commute and do not contribute to $g_k$, so that the expectation value $g_k$ can be computed as follows.
\begin{eqnarray}
\begin{aligned}
g_k &= \langle \psi(0) |U^{\dagger}(H_f,\gamma)  U^{\dagger} (H_B,\beta) \sigma_z^{(k)} U(H_B,\beta) U(H_f,\gamma)  |\psi(0)\rangle \\
&= \langle+^3| U^{\dagger}(A_{k',k},\gamma)  U^{\dagger}(A_{k,k''},\gamma)   \sigma_z^{(k)}  U^{\dagger} (b_k,\beta)\sigma_x^{k}  U(b_k,\beta_1) U(A_{k,k''},\gamma)  U^{\dagger}(A_{k',k},\gamma)  \sigma_z^{(k)} |+^3\rangle,
\end{aligned}
\end{eqnarray}
where $k' < k < k''$.

\bibliographystyle{IEEEtran}
%%\addbibresource{/users/jingjing.cui/dropbox/EndnoteLib/myrefv1}
%%  \linespread{1.1}\selectfont
%%\bibliography{C:/Users/jc12n19/Dropbox/EndnoteLib/myrefv1}
\bibliography{/users/jingjing.cui/dropbox/EndnoteLib/myrefv1}

\end{document}